\documentclass[journal,10pt]{IEEEtran}

\usepackage{amsmath,amssymb, multicol} 
\usepackage{graphicx}
\usepackage{epsfig}
\usepackage{epstopdf}
\usepackage{cite}
\DeclareMathAlphabet{\mathcal}{OMS}{cmsy}{m}{n}
\usepackage{times}
\usepackage{setspace} 
\usepackage{color}
\usepackage[justification=centering]{caption}
\usepackage{booktabs}
\usepackage{algorithm}
\usepackage{algpseudocode}
\usepackage{amsmath}
\usepackage{makecell}
\usepackage{caption}
\newcommand{\larrow}{\overset{\scriptscriptstyle\leftarrow}}
\newcommand{\rarrow}{\overset{\scriptscriptstyle\rightarrow}}
\usepackage{stfloats}
\makeatletter
\renewcommand{\maketag@@@}[1]{\hbox{\m@th\normalsize\normalfont#1}}%
\makeatother
\newenvironment{salign}{\par\nobreak\small\noindent\align}{\endalign}

\begin{document}
\title{Bayesian Receiver Design for Grant-Free NOMA with Message Passing Based Structured Signal Estimation  }

\author{Yuanyuan Zhang, Zhengdao Yuan, Qinghua Guo, \textit{Senior Member}, \textit{IEEE}, Zhongyong Wang, Jiangtao Xi, \textit{Senior Member}, \textit{IEEE}, and Yonghui Li, \textit{Fellow}, \textit{IEEE}
	


\thanks{ Y. Zhang  is with the School of Information Engineering, Zhengzhou University, Zhengzhou 450001, China, and the School of Electrical, Computer and Telecommunications Engineering, University of Wollongong, Wollongong, NSW 2522, Australia (e-mail: ieyyzhang@gmail.com). }
\thanks{ Z. Yuan is with the Artificial Intelligence Technology Engineering Research Center, Henan Redio \& TV University, and the School of Information Engineering, Zhengzhou University, Zhengzhou 450002, China. He was also with the School of Electrical, Computer and Telecommunications Engineering, University of Wollongong, Wollongong, NSW 2522, Australia (e-mail: yuan\_zhengdao@foxmail.com). }
\thanks{Q. Guo and J. Xi are with the School of Electrical, Computer and Telecommunications Engineering, University of Wollongong, Wollongong, NSW 2522, Australia (e-mail: qguo@uow.edu.au, jiangtao@uow.edu.au).}
\thanks{ Z. Wang  is with the School of Information Engineering, Zhengzhou University, Zhengzhou 450001, China (e-mail: zywangzzu@gmail.com ). }
\thanks{ Y. Li is with the School of Electrical and Information Engineering, University of Sydney, Sydney, NSW 2006, Australia (e-mail: yonghui.li@sydney.edu.au).}
}
\maketitle

\begin{abstract}
Grant-free non-orthogonal multiple access (NOMA) is promising to achieve low latency massive access in Internet of Things (IoT) applications. In grant-free NOMA, pilot signals are often used for user activity detection (UAD) and channel estimation (CE) prior to multiuser detection (MUD) of active users. However, the pilot overhead makes the communications inefficient for IoT devices with sporadic transmissions and short data packets, or when the channel coherence time is short. Hence, it is desirable to improve the efficiency by avoiding the use of pilot signals, which can also further achieve lower latency.
This work focuses on  Bayesian receiver design for grant-free low density signature orthogonal frequency division multiplexing (LDS-OFDM), where each user is allocated a unique low density spreading sequence.  We propose to use the low density spreading sequences for active user detection, thereby avoiding the use of pilot signals. Firstly, the task of joint UAD, CE and MUD is formulated as a structured signal estimation problem. Then message passing   based Bayesian approach is developed  to solve the structured signal estimation problem. In particular, belief propagation (BP), expectation propagation (EP) and mean field (MF) message passing are used to develop efficient hybrid message passing algorithms to achieve trade-off between performance and complexity. Simulation results  demonstrate the effectiveness of the proposed receiver for grant-free LDS-OFDM without the use of pilot signals.
\end{abstract}

\begin{IEEEkeywords}
 Grant-free, non-orthogonal multiple access (NOMA), multiuser detection, message passing, Bayesian inference.
\end{IEEEkeywords}

\IEEEpeerreviewmaketitle

\section{Introduction}
\IEEEPARstart{T}{he} explosion of small and cheap machine-type devices with sensing and communication capability is paving the way towards smart home, smart city, smart health care and factory automation \cite{IOT2015}.
As one of the major application scenarios in the fifth generation (5G) wireless communications, massive machine type communications (mMTC) aims to accommodate  massive connections and sporadic short-burst transmissions in  Internet of Things (IoT) systems \cite{notOMP2014,5Gscenarios2016,IOT2016}.
Due to the limited spectral resource, the conventional orthogonal multiple access (OMA) techniques cannot meet the demands on massive connections in massive IoT networks. Non-orthogonal multiple access (NOMA), where a resource block can be used to serve multiple users, is considered as a promising technology to support mMTC\cite{NOMA2013,daiNOMA2015,MassiveNOMA2017,NOMA2018}.
In addition, the conventional grant-based access protocols  with handshaking procedure may lead to excessive overhead, long and uncertain  latency, which can be unacceptable in sporadic short-burst IoT traffic as the communication becomes inefficient due to the small amount of payload data \cite{MTC2014, mMTC2016}.
Therefore, grant-free access without handshaking procedure is
highly desirable, where users can   transmit data at any
time slot and active users have to be identified by the access point before data detection.


The major tasks for the access point  in grant-free NOMA system includes user activity detection (UAD), channel estimation (CE) and multiuser detection (MUD) of active users. In existing works,   UAD is coupled with  CE and/or MUD, e.g., CE is followed by joint UAD and MUD \cite{AMP2017,wang2016Dcs,du2017Dcs,M2M2017MUD, xin2019,Bichai2015CS}, pilot assisted joint UAD and CE \cite{chen2018sparse, BSBL2018,UADandCE2018,Fu2019UADandCE} is followed by MUD \cite{Novel2008,lowMPA2012low,receiverMPA2012}, and UAD, CE and MUD are performed jointly with the aid of pilot signals \cite{du2018joint,WenChen2017Proc,WenChen2018TWC}.
By exploiting that  only a small fraction of the  users in the network are active at a time, i.e., the distribution of active users is sparse, the problem of joint UAD and MUD or joint UAD and CE  has been formulated under the compressive sensing (CS) framework \cite{CS2006,MUDbasedCS2012,SparseSP2018}.
In \cite{AMP2017,wang2016Dcs,du2017Dcs,M2M2017MUD,xin2019,Bichai2015CS}, with the assumption that the channel state information (CSI) is available to the receiver, various approaches such as those based on approximate message passing (AMP)\cite{AMP2017}, orthogonal matching pursuit  (OMP)\cite{wang2016Dcs} and prior-information aided adaptive subspace pursuit (PIA-ASP) \cite{du2017Dcs} were developed  for joint UAD and MUD.   However, in many  scenarios, particularly dynamic ones such as Internet of vehicles (IoV), the CSI varies over time and has to be estimated frequently.
With the assist of pilot signals,  joint UAD, CE and MUD are performed jointly in \cite{du2018joint,WenChen2017Proc,WenChen2018TWC}, and joint UAD and CE are carried out in \cite{chen2018sparse, BSBL2018,UADandCE2018,Fu2019UADandCE} to identify active users and estimate their CSI (which are then used for MUD).
However, the use of pilot  introduces excessive overhead, making communications inefficient for IoT devices with sporadic transmission and short data packets, or when the channel coherence time is short. Hence, it is desirable to improve the efficiency by avoiding the use of pilot signals, which can also further achieve lower latency.

In this work, we consider the NOMA scheme low density signature orthogonal frequency division multiplexing (LDS-OFDM) \cite{receiverMPA2012} and investigate the receiver design  for grant-free LDS-OFDM, where pilot signals are not used. In LDS-OFDM, as each user is allocated a unique low density spreading (LDS) sequence, it is possible to identify active users based on the LDS sequences, therefore avoiding the use of pilot signals.
We first show that joint UAD, CE and MUD without the use of pilot can be formulated as an interesting structured signal estimation problem, where the structures of the signals are brought by the LDS matrix of LDS-OFDM and the discreteness of  transmitted signals.  The structures can be fully exploited to identify active users and estimate active users' channel gain and their symbols, so pilot signals are not necessary.
We then consider Bayesian approaches to the structured signal estimation problem, and develop efficient message passing  based Bayesian inference algorithms which ran in a graph representation of the system.
Specifically, belief propagation (BP)  \cite{FactorG2001} and expectation propagation (EP) \cite{EP2001,BPEP2007} are combined for UAD and CE, and mean field (MF) based message passing  \cite{MF2002,VMP2005} is used at observation factors for noise power estimation and MUD. It is noted that  UAD, CE and MUD are seamlessly integrated based on their message passing implementation.
By introducing some auxiliary variables to break down the observation factors, we further develop another hybrid message passing algorithm, where BP and MF are merged to improve the system performance substantially. Extensive simulation results are provided to demonstrate the effectiveness of the proposed algorithms.

The rest of this paper is organized as follows. Section II describes the system model and the problem formulation. Message passing based Bayesian receiver is developed in Section III. Numerical simulation results are provided in Section IV, followed by conclusions in Section V.

\textit{Notation}- Lowercase and uppercase letters denote scalars. Boldface lowercase and uppercase letters denote column vectors and matrices, respectively. The superscriptions $(\cdot)^T $ and $(\cdot)^H $ denote the transpose and conjugate transpose operations, respectively, and $\varpropto $ denotes equality of functions up to a scale factor. The functions $  \mathcal{CN}(x;\hat{x},\sigma^2_x) $  stands for  a complex Gaussian distribution with mean $\hat{x} $ and variance $\sigma^2_x $. As the convention, {\small $ \left< f\left( x,y,z\right) \right>_{f\left(y\right)f\left(z\right)}\text{=}\iint \!\! {f( x,y,z)f(y)f(z)dydz} $}  is used to denote the marginalization operator. The expectation operator with respect to a probability density function (PDF) $ g(x) $ is expressed by {\small $\left<x\right>_{g\left(x\right)}\text{=}\int {xg(x)dx}/\int {g({{x}'})d{{x}'}} $}, and {\small $ \text{Var}\left[ x \right]_{g(x)} \text{=} \left< |x|^2 \right>_{g(x)} - |\left< x \right>_{g(x)}|^2 $} stands for the variance of $ x $.

\section{ System Model and Problem Formulation}

\begin{figure}[!t]
	\centering
	\includegraphics[width=3.5in]{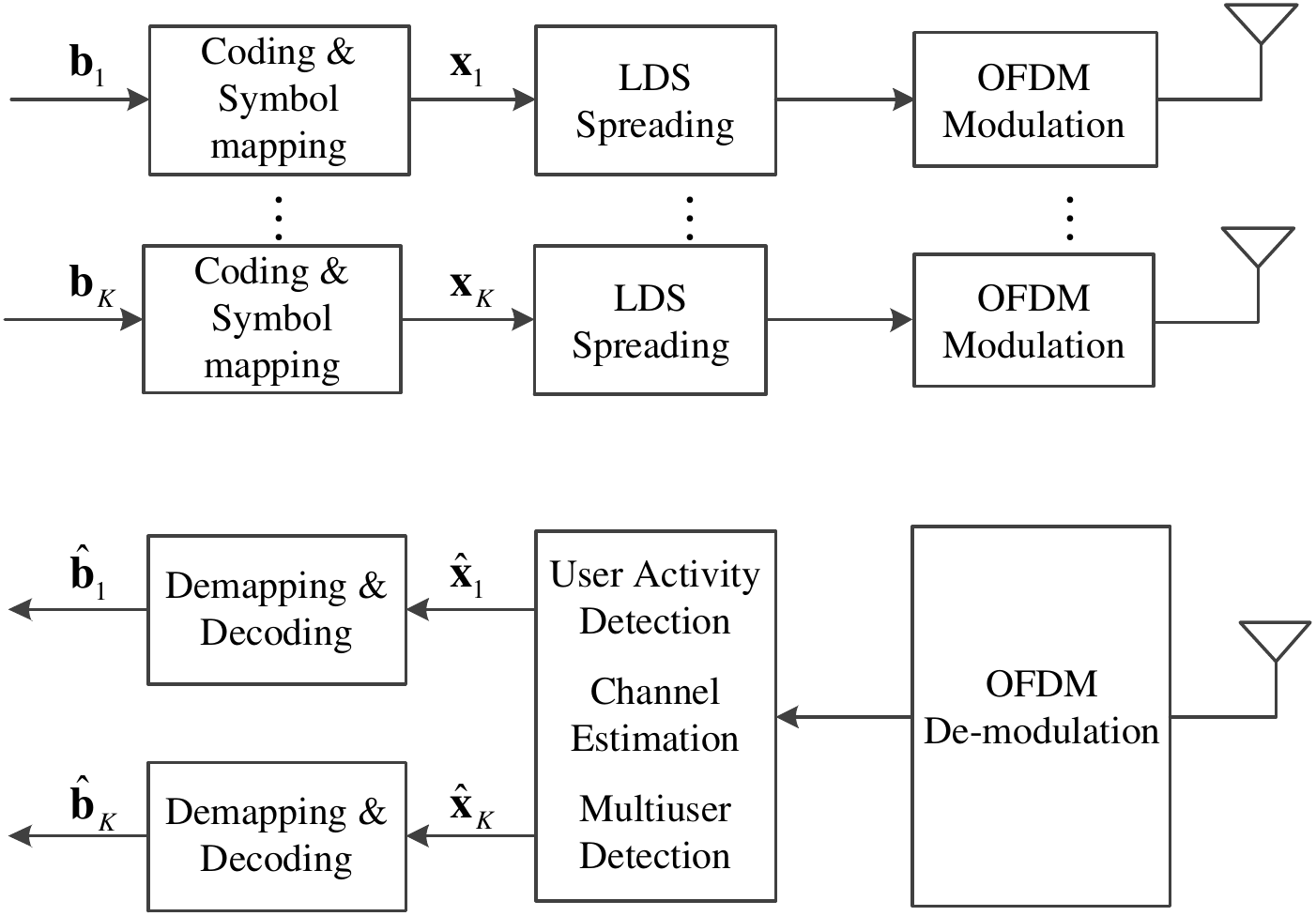}
	\caption{Block diagram of LDS-OFDM with $ K $ active users}\label{fig:BD}
\end{figure}

As show in Fig. \ref{fig:BD}, we assume an uplink LDS-OFDM system with   $ N $ subcarriers and $ U $  users, where the number of active users denoted by $ K $ can be much smaller than  $ U $.
The input bit stream $ \textbf{b}_k $ of user $ k $ is coded and mapped to a symbol sequence $ \textbf{x}_k \in \mathbb{C}^{ L \times 1}  $, where $ L $ is the length of the  sequence. Then, each symbol in $ \textbf{x}_k $ is spread onto $ N $ OFDM subcarriers  using a unique low-density spreading sequence $ \textbf{s}_k $ of length $ N $, where $ \textbf{s}_k $ is a column of the  LDS matrix $ \textbf{S} =[ \textbf{s}_1, \textbf{s}_2, \cdots,\textbf{s}_{U}] $.
For simplicity, we consider the LDS matrix $ \textbf{S}$ with a regular structure \cite{Novel2008}, where $ \textbf{S}$ has the same number of non-zero elements in each column denoted by $ d_c $, and also the same number of non-zero elements in each row denoted by $ d_r $, i.e., each users occupies  $ d_c $ subcarriers, and each subcarrier accommodates  $ d_r $ users. It is noted that the extension of this work  to the case of irregular LDS structure is straightforward.
As only $ K $  out of $ U $  users are active, we use $ {z}_{\scriptscriptstyle k} \in \mathcal{U} \triangleq \{ 1,2,\cdots, U \}$ to denote the identity of active user $ k $. Hence, we can define an active user LDS matrix $\textbf{S}_a = [\textbf{s}_{z_1}, \textbf{s}_{z_2}, \cdots,\textbf{s}_{z_{\scriptscriptstyle K}}]$ with size $ N\times K $, i.e., $ \textbf{S}_a $ is a sub-matrix of $ \textbf{S} $.  We assume that the channel is approximately static within a data block. The received signal on the $ n $-th subcarrier is the superposition of the signals of $ K $ active users, i.e.,
\begin{equation}
\textbf{y}_n =\sum_{k=1}^{K}g_{n,k}s_{n,z_k}\textbf{x}_{k} + \textbf{w}_n, n=1,2,\cdots,N,
\end{equation}
where $ \textbf{y}_n \in \mathbb{C}^{L\times 1} $, $ g_{n,k} $ is the channel gain of active user $ k $ on the $ n $-th subcarrier,  $ s_{n,z_k} $ is the $ n $-th component of the spreading sequence  $ \textbf{s}_{z_k} $, which is non-zero  if active user $ k $ transmits signal on the $ n $-th subcarrier and 0 otherwise, and the noise vector on the $ n $-th subcarrier  $ \textbf{w}_n \sim   \mathcal{CN}(\textbf{w}_n;0,\sigma^2_w\textbf{I}_{L})$. In a matrix form, the received signals over $ N $ subcarriers can be expressed as
\begin{equation} \label{eq-Y}
\textbf{Y}=\textbf{HX} + \textbf{W} ,
\end{equation}
where $ \textbf{Y}= [\textbf{y}_1  , \textbf{y}_2  , \cdots , \textbf{y}_{N} ]^T $ is an $ N\times L $ matrix,
\begin{equation}\label{eq-H}
\textbf{H}= \textbf{G} \odot  \textbf{S}_a
\end{equation}
($ \odot $ represents the Hadamard product ) is an  $ N\times K $ equivalent channel matrix, whose $ (n,k) $-th element $ h_{n,k} =g_{n,k}s_{n,z_k}$, $ \textbf{X}=[\textbf{x}_1 ,\textbf{x}_2 ,\cdots,\textbf{x}_{\scriptscriptstyle K} ]^T $ is the transmitted symbol matrix of size $ K\times L $, and $ \textbf{W} \in \mathbb{C}^{\scriptscriptstyle N\times L} $ is the noise matrix.

Our objective is to estimate $ \textbf{H} $ and $ \textbf{X} $ simultaneously based on $ \textbf{Y} $. This is possible thanks to
the structures of  $ \textbf{H} $ (brought by the active user LDS matrix $ \textbf{S}_a $) and $ \textbf{X}$ (brought by the
discreteness of transmitted signals), i.e.,  the columns of $  \textbf{S}_a $ in (\ref{eq-H}) are randomly drawn from those of the LDS matrix $\textbf{S} $, and the entries of $ \textbf{X} $ are discrete values  (drawn from a constellation set), which are mapped from coded bits. These structures can be exploited to recover  $ \textbf{H} $ and $ \textbf{X}$. However, when the constellation is symmetric, the solution may not be unique due to the phase ambiguity. The problem can be overcome by using  rotationally-invariant coded modulation schemes, e.g., the rotationally-invariant trellis coded   (RI-TCM) encoder in \cite{90RIRCM1999}, which combines the operations of coding and modulation.

\section{ Bayesian receiver design with message passing based structured signal estimation }
To  jointly estimate $ \textbf{H} $ and $ \textbf{X}$  in (\ref{eq-Y}) from the received signal $ \textbf{Y} $ by exploiting the structures of $ \textbf{H} $ and $ \textbf{X}$, graphic model and message passing based Bayesian inference is investigated in this section.  In addition, the computational complexity is analysed.

\subsection{Factor graph representation}
\begin{figure}
	\centering
	\includegraphics[width=3.5in]{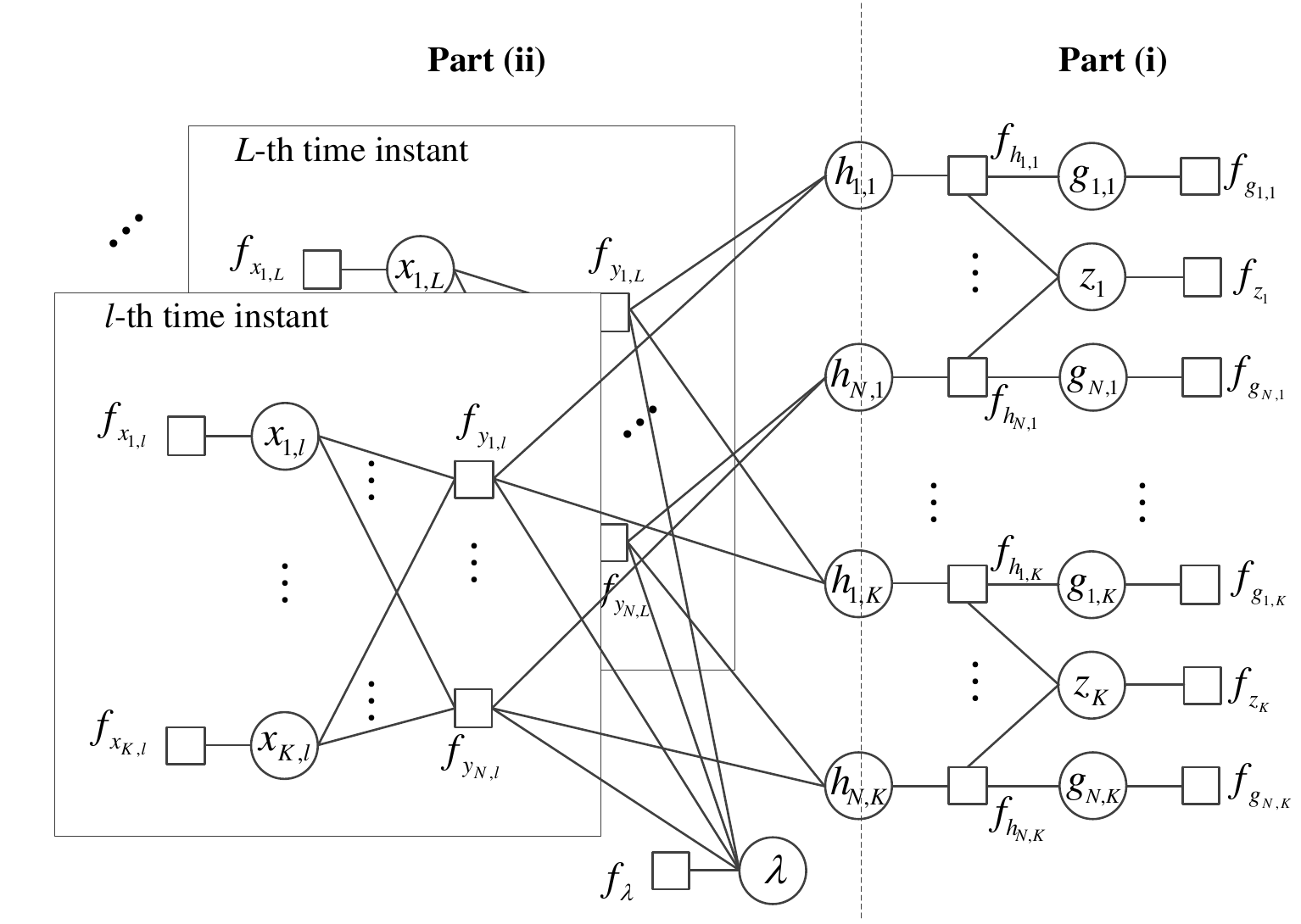}
	\caption{Factor graph representation of (\ref{eq:pdf}). }\label{FG}
\end{figure}

We note that ${z}_{\scriptscriptstyle k} \in \mathcal{U} \triangleq \{ 1,2,\cdots, U \}$, which indicates that the active user  $ k $ employs the $ {z}_{\scriptscriptstyle k} $-th spreading sequence $ \textbf{s}_{z_k} $. We assume the a priori PDF of  ${z}_{\scriptscriptstyle k}$ is {\small $p({z}_{\scriptscriptstyle k})= \sum_{u\in \mathsf{\mathcal{U}}} \frac{1}{U} \delta({z}_{\scriptscriptstyle k}-u )$}.
As shown in (\ref{eq-H}), the equivalent channel matrix $ \textbf{H} $ is the Hadamard product of the channel gain matrix $ \textbf{G} $ and the active user LDS matrix $ \textbf{S}_a$, which are independent of each other.  Hence, we have
\begin{equation}
\begin{split}
 p( \mathbf{H,G,z})
&= p( \mathbf{H}| \mathbf{G,z}) p( \mathbf{G}) p(\mathbf{z}) \\
& =  \prod\limits_{k=1}^{K} { \left( p\left( {{z}_{\scriptscriptstyle k}} \right) \prod\limits_{n=1}^{N}{p\left( {h_{n,k}}\left| g_{n,k},{{z}_{\scriptscriptstyle k}} \right. \right)p\left( g_{n,k} \right)} \right)},
\end{split}
\end{equation}
where $ \textbf{z} = [{z}_{1}, {z}_{2},\cdots, {z}_{\scriptscriptstyle K}] $, $ p({h_{n,k}}| {g_{n,k}},{{z}_{\scriptscriptstyle k}}) $ represents a hard constraint, i.e., $ \delta (h_{n,k}- g_{n,k}s_{n,{z}_{\scriptscriptstyle k}}) $, and $ p({g_{n,k}}) =   \mathcal{CN}({g_{n,k}}; 0, 1) $. We assume that the noise precision $ \lambda=1/\sigma^2_w $  is unknown, and it has a prior $ p(\lambda) \propto 1/ \lambda $. Given $ \textbf{Y} $, the joint a posteriori PDF of $ \textbf{X, H, G, z} $ and $ \lambda $  can be expressed as
\begin{small}
	\begin{align} \label{eq:pdf}
	p( \mathbf{X,H,G,z},\lambda | \mathbf{Y} )
	\notag & \varpropto  p( \mathbf{Y}| \mathbf{X,H},\lambda ) p( \mathbf{X} )  p( \mathbf{H, G,z}) p( \lambda)\\
	\notag & =p(\lambda)\prod\limits_{l=1}^{L}{\left( \prod\limits_{k=1}^{K}{p(x_{\scriptscriptstyle k,l})\prod\limits_{n=1}^{N}{p( y_{n,l} | \lambda ,{h_{n,k}},x_{\scriptscriptstyle k,l},\forall k)}} \right)} \\
	& \quad \cdot \prod\limits_{k=1}^{K} { \left( p\left( {{z}_{\scriptscriptstyle k}} \right) \prod\limits_{n=1}^{N}{p\left( {h_{n,k}}\left| {g_{n,k}},{{z}_{\scriptscriptstyle k}} \right. \right)p\left( {g_{n,k}} \right)} \right)} ,
	\end{align}
\end{small}\normalsize
where {\small $ p(x_{\scriptscriptstyle k,l})=\sum_{q\in \mathsf{\mathcal{X}}}{\frac{1}{Q}\delta ( x_{\scriptscriptstyle k,l}-q )} $} and {\small $ p( y_{n,l} | \lambda ,{h_{n,k}},x_{\scriptscriptstyle k,l},\forall k) =  \mathcal{CN}(y_{n,l}; \sum_{k=1}^K h_{n,k}x_{\scriptscriptstyle k,l}, 1/\lambda)$}.
To facilitate the factor graph representation of the factorization in (\ref{eq:pdf}),  we introduce the notations in Table \ref{tab1}, showing the correspondence between the factor labels and the underlying PDFs they represent. The factor graph representation of (\ref{eq:pdf}) is shown in Fig. \ref{FG}.

We divide the factor graph in Fig. \ref{FG} into two parts labeled by Part (i) and Part (ii).
As we can see Part (i) represents the structure of $ \textbf{H} $, where
message passing rules are derived by combining BP and EP, this part functions as active user detector and realizes channel estimation in conjunction with Part (ii). Part (ii) functions as multiuser detector and noise estimator, where the structure of $ \textbf{X} $ is included. In this part, two massage passing algorithms are developed based on the pure MF and hybrid BP-MF, respectively, which can achieve  different trade-off between  computational complexity and performance.


\begin{table}[t]
	\renewcommand\arraystretch{1.2} 
	\centering
	\caption{The factors involved in the factorization in (\ref{eq:pdf})}\label{tab1}
	\resizebox{8cm}{2cm}{
		\begin{tabular}{l l l}
			\toprule[1pt]
			\textbf{Factor} & \textbf{Distribution} & \textbf{Functional Form} \\
			\midrule[0.5pt]
			$ f_\lambda(\lambda)$ & $ p(\lambda) $ & $ 1/\lambda $ \\
			$ f_{x_{\scriptscriptstyle k,l}}( x_{\scriptscriptstyle k,l}) $ & $ p( x_{\scriptscriptstyle k,l})$ & $ \sum\limits_{q\in \mathsf{\mathcal{X}}}{\frac{1}{Q}\delta ( x_{\scriptscriptstyle k,l}-q )} $ \\
			$ f_{y_{n,l}}(  \lambda ,{h_{n,k}},x_{\scriptscriptstyle k,l},\forall k)$ & $ p( y_{n,l} | \lambda ,{h_{n,k}},x_{\scriptscriptstyle k,l},\forall k  ) $ & $   \mathcal{CN}(y_{n,l}; \sum\limits_{k=1}^K h_{n,k}x_{\scriptscriptstyle k,l}, 1/\lambda) $\\
			$f_{h_{n,k}} ({h_{n,k}}, {g_{n,k}}, {z}_{\scriptscriptstyle k} )$  & $ p ({h_{n,k}}| {g_{n,k}},{{z}_{\scriptscriptstyle k}}) $ & $ \delta (h_{n,k}- g_{n,k}s_{n,{z}_{\scriptscriptstyle k}}) $\\
			$ f_{g_{n,k}} ({g_{n,k}})$ & $ p({g_{n,k}}) $ & $   \mathcal{CN}({g_{n,k}}; 0, 1) $ \\
			$ f_{{z}_{\scriptscriptstyle k}} ({z}_{\scriptscriptstyle k}) $ & $ p({z}_{\scriptscriptstyle k}) $ & $ \sum \limits_{u\in \mathsf{\mathcal{U}}}\frac{1}{U}\delta({z}_{\scriptscriptstyle k}-u ) $ \\
			\bottomrule[1pt]
	\end{tabular}}
\end{table}

In the following, we detail the forward (from left to right) and backward (from right to left)  message computations at each node of Part (i) and Part (ii), and some approximations are introduced to reduce the computational complexity. We use $ I_{A\to B}(x) $ to denote a message passed from a variable node (function node) $ A $ to a function node (variable node) $ B $, which is a function of $ x $.  The notations $ m $ and $ v $ are used to denote the mean and variance of a Gaussian message specified by their subscripts. The arrows over $ m $ and   $ v $ represent the directions of Gaussian massage passing.
Note that, if a forward message computation requires backward  messages, we use the messages in previous iteration by default.

\begin{figure*}[b]
	\begin{small}
	\begin{align*}\label{eq:BetaKU}
	\beta _{k}^{u}
	= \frac{b\left( {{z}_{\scriptscriptstyle k}}=u \right)}{\sum\limits_{{u}'\in \mathsf{\mathcal{U}}}{b\left( {{z}_{\scriptscriptstyle k}}={u}' \right)}}
	= \frac{\prod\limits_{n=1}^{N}{\left[ \left( 1-{{s}_{n,u}} \right)\mathcal{CN}\left( 0;{{{\rarrow{m}}}_{{h_{n,k}}}},{{{\rarrow{v}}}_{{h_{n,k}}}} \right)+{{s}_{n,u}}\mathcal{CN}\left( {{{\rarrow{m}}}_{{h_{n,k}}}};{{{\larrow{m}}}_{{g_{n,k}}}},{{{\rarrow{v}}}_{{h_{n,k}}}}+{{{\larrow{v}}}_{{g_{n,k}}}} \right) \right]}}{\sum\limits_{{u}'\in \mathsf{\mathcal{U}}}{\left( \prod\limits_{n=1}^{N}{\left[ \left( 1-{{s}_{n,{u}'}} \right)\mathcal{CN}\left( 0;{{{\rarrow{m}}}_{{h_{n,k}}}},{{{\rarrow{v}}}_{{h_{n,k}}}} \right)+{{s}_{n,{u}'}}\mathcal{CN}\left( {{{\rarrow{m}}}_{{h_{n,k}}}};{{{\larrow{m}}}_{{g_{n,k}}}},{{{\rarrow{v}}}_{{h_{n,k}}}}+{{{\larrow{v}}}_{{g_{n,k}}}} \right) \right]} \right)}} \tag{8}
	\end{align*}
	\begin{align*} \label{eq:gamma}
	\gamma _{k}^{u}=\frac{\prod\limits_{{n}'\ne n}{\left[ \left( 1-{{s}_{{n}',u}} \right)\mathcal{CN}\left( 0;{{{\rarrow{m}}}_{{{h}_{{n}',k}}}},{{{\rarrow{v}}}_{{{h}_{{n}',k}}}} \right)+{{s}_{{n}',u}}\mathcal{CN}\left( {{{\rarrow{m}}}_{{{h}_{{n}',k}}}};{{{\larrow{m}}}_{{{g}_{{n}',k}}}},{{{\rarrow{v}}}_{{{h}_{{n}',k}}}}+{{{\larrow{v}}}_{{{g}_{{n}',k}}}} \right) \right]}}{\sum\limits_{{u}'\in \mathsf{\mathcal{U}}}{\left( \prod\limits_{{n}'\ne n}{\left[ \left( 1-{{s}_{n',{u}'}} \right)\mathcal{CN}\left( 0;{{{\rarrow{m}}}_{{h_{n',k}}}},{{{\rarrow{v}}}_{{h_{n',k}}}} \right)+{{s}_{n',{u}'}}\mathcal{CN}\left( {{{\rarrow{m}}}_{{h_{n',k}}}};{{{\larrow{m}}}_{{g_{n',k}}}},{{{\rarrow{v}}}_{{h_{n',k}}}}+{{{\larrow{v}}}_{{g_{n',k}}}} \right) \right]} \right)}} \approx \beta _{k}^{u} \tag{10}
	\end{align*}
	\end{small}	
\end{figure*}

\subsection{Message Passing in Part (i)}
With the output forward message from Part (ii), BP and EP based message passing is used to realize the estimation of  {\small $\{ {z}_{\scriptscriptstyle k}, \forall k \}$} and {\small $\{ h_{n,k}, \forall n,k \}$}, where the function nodes {\small $\{f_{h_{n,k}} , \forall n,k \}$} are handled by the BP rule, and some messages are approximated to be Gaussian with EP to reduce the computational complexity.

\begin{spacing}{1.1}
With the message  {\small $I_{g_{n,k} \rightarrow f_{h_{n,k}}}\big(g_{n,k}\big) \varpropto  \mathcal{CN}  \big(g_{n,k} ; {{{\larrow{m}}}_{{g_{n,k}}}}, {{{\larrow{v}}}_{{g_{n,k}}}} \big)$}  and
{\small $ {{I}_{{h_{n,k}}\to{{f}_{{h_{n,k}}}}}} \big({h_{n,k}}\big)  \varpropto \! \mathcal{CN} \big( {h_{n,k}};{{{\rarrow{m}}}_{{h_{n,k}}}},{{{\rarrow{v}}}_{{h_{n,k}}}} \big)$} output from Part (ii) (the calculation of {\small $ {{I}_{{h_{n,k}}\to{{f}_{{h_{n,k}}}}}} \big({h_{n,k}}\big)$} is relate to (\ref{eq:htofh1}) based on MF or (\ref{eq:htofh2}) based on BP-MF), the forward message {\small $I_{f_{h_{n,k}} \rightarrow z_{k}}(z_{k})$} is given by
\end{spacing} \vspace{-4mm}
\begin{small}
\begin{align}
{{I}_{{{f}_{{h_{n,k}}}}\to {{z}_{\scriptscriptstyle k}}}}\left( {{z}_{\scriptscriptstyle k}} \right)
\notag & = {\Big\langle  {{f}_{{h_{n,k}}}}\left( {h_{n,k}},{g_{n,k}},{{z}_{\scriptscriptstyle k}} \right)  \Big\rangle }_{ {{I}_{{h_{n,k}}\to {{f}_{{h_{n,k}}}}}} ({h_{n,k}}) {{I}_{{g_{n,k}}\to {{f}_{{h_{n,k}}}}}} ({g_{n,k}})} \\
\notag & = \left( 1-{{s}_{n,{{z}_{\scriptscriptstyle k}}}} \right)\mathcal{CN}\left( 0;{{{\rarrow{m}}}_{{h_{n,k}}}},{{{\rarrow{v}}}_{{h_{n,k}}}} \right) \\
& \quad \ +{{s}_{n,{{z}_{\scriptscriptstyle k}}}}\mathcal{CN}\left( {{{\rarrow{m}}}_{{h_{n,k}}}};{{{\larrow{m}}}_{{g_{n,k}}}},{{{\rarrow{v}}}_{{h_{n,k}}}}+{{{\larrow{v}}}_{{g_{n,k}}}} \right).
\end{align}
\end{small}\normalsize

Then, the belief of $ {z}_{\scriptscriptstyle k} $ can be updated as
\begin{align}
\notag b\left(z_{k}\right) & \varpropto \prod_{n=1}^{N} I_{f_{h_{n,k}} \rightarrow z_{k}}\left(z_{k}\right) I_{f_{z_{k}} \rightarrow z_{k}}\left(z_{k}\right)\\
& \varpropto \sum\limits_{u\in \mathsf{\mathcal{U}}}{\beta _{k}^{u}\delta \left( {{z}_{\scriptscriptstyle k}}-u \right)} ,
\end{align}
where $ \beta _{k}^{u} $ is calculated by (\ref{eq:BetaKU}) at the bottom of the page. Note that $ b\left(z_{k}\right) $ is used to determine the identity of active user $ k $, i.e., $ \hat z_{k} = \mathop{\arg\max}_{z_{k}} b\left(z_{k}\right)$ indicates that active user $ k $ employs the LDS sequence $ \textbf{s}_{\hat z_{k}} $, so active user $ k $ is identified.
The backward message {\small $I_{{{z}_{\scriptscriptstyle k}}\to {{f}_{{h_{n,k}}}}} \left( {{z}_{\scriptscriptstyle k}} \right) $} can be expressed as
\begin{small}
\begin{align*}
{{I}_{{{z}_{\scriptscriptstyle k}}\to {{f}_{{h_{n,k}}}}}}\left( {{z}_{\scriptscriptstyle k}} \right)
\notag &= \prod_{n'\neq n} I_{f_{h_{n', k}} \rightarrow z_{k}}\left(z_{k}\right) I_{f_{z_{k}} \rightarrow z_{k}}\left(z_{k}\right)\\
&\varpropto \sum\limits_{u\in \mathsf{\mathcal{U}}}{\gamma _{k}^{u}\delta \left( {{z}_{\scriptscriptstyle k}}-u \right)} \tag{9},
\end{align*}
\end{small}\normalsize
where $ \gamma _{k}^{u} $ is calculated by (\ref{eq:gamma}) at the bottom of the page and it can be approximated as $ \beta _{k}^{u} $ to reduce the computational complexity.
Thus, the backward message {\small $I_{{{f}_{{h_{n,k}}}}\to {h_{n,k}}} \left( {h_{n,k}} \right)$}  is given by
\begin{salign}
\setcounter{equation}{10}
\notag & I_{{{f}_{{h_{n,k}}}}\to {h_{n,k}}} \left( {h_{n,k}} \right)
= {\Big\langle {{f}_{{h_{n,k}}}} ( {h_{n,k}},{g_{n,k}},{{z}_{\scriptscriptstyle k}})   \Big\rangle}_{{{I}_{{{z}_{\scriptscriptstyle k}}\to {{f}_{{h_{n,k}}}}}} {{I}_{{g_{n,k}}\to {{f}_{{h_{n,k}}}}}}}\\
& = \sum\limits_{u\in \mathsf{\mathcal{U}}}{\beta _{k}^{u}}\left[ \left( 1-{{s}_{n,u}} \right)\delta \left( {h_{n,k}} \right)+{{s}_{n,u}}\mathcal{CN}\left( {h_{n,k}};{{{\larrow{m}}}_{{g_{n,k}}}},{{{\larrow{v}}}_{{g_{n,k}}}} \right) \right],
\end{salign}
which is not  Gaussian, so that the belief {\small $  b\left({h_{n,k}}\right) \varpropto I_{{{f}_{{h_{n,k}}}}\! \to  {h_{n,k}}} \left({h_{n,k}}\right){{I}_{{h_{n,k}}  \to \! {{f}_{{h_{n,k}}}}}} \left({h_{n,k}}\right) $} is not Gaussian either.  It is difficult to calculate the backward message {\small $I_{{h_{n,k}}\to f_{\psi _{\scriptscriptstyle n,k,l}}} \left( {h_{n,k}} \right)$} in (\ref{eq:h2fpsi}). We propose to use the EP method to overcome this problem. We first approximate   {\small $  b\left({h_{n,k}}\right)$} to be Gaussian, i.e.,
\begin{salign}\label{eq:blfH}
\notag  b\left( {h_{n,k}} \right) &\approx b^{\scriptscriptstyle G}\left( {h_{n,k}} \right) \\
\notag & = \text{Proj}_{\scriptscriptstyle G}\left\{ I_{{{f}_{{h_{n,k}}}}\to {h_{n,k}}}^{{}}\left( {h_{n,k}} \right){{I}_{{h_{n,k}}\to {{f}_{{h_{n,k}}}}}}\left( {h_{n,k}} \right) \right\} \\
& \triangleq \mathcal{CN}\left( {h_{n,k}}; {{{\hat{h}}}_{n,k}},v_{{h_{n,k}}} \right),
\end{salign}
where $ \text{Proj}_{\scriptscriptstyle G}\{*\} $ denotes the operation of Gaussian approximation, and the mean and the variance of $  {h_{n,k}} $ can be calculated by moment matching, i.e.,
\begin{salign}\label{eq:epMh}
{{{\hat{h}}}_{n,k}}
\notag & ={{\Big\langle h_{n,k}  \Big\rangle }_{{{b}}({h_{n,k}})}} \\
&=\frac{1}{ O_{h_{n,k}}} \sum\limits_{u\in \mathsf{\mathcal{U}}}  {\beta _{k}^{u}}{{s}_{n,u}}{{{ {m}}}_{n,k}}\mathcal{CN} \big( {{{\rarrow{m}}}_{{h_{n,k}}}};{{{\larrow{m}}}_{{g_{n,k}}}},{{{\rarrow{v}}}_{{h_{n,k}}}}+{{{\larrow{v}}}_{{g_{n,k}}}}\big),
\end{salign}
\begin{salign}\label{eq:epVh}
 v_{{h_{n,k}}} \notag & ={{\text{Var}\Big[ h_{n,k}  \Big] }_{{{b}}({h_{n,k}})}}\\
&=\! \frac{1}{O_{h_{n,k}}}   \sum\limits_{u\in \mathsf{\mathcal{U}}}  \! {\beta _{k}^{u}}{{s}_{n,u}}  \! \left[ |{m}_{n,k}|^{2}\!+\!{{{ {v}}}_{n,k}}\right] \mathcal{CN} \big( {{{\rarrow{m}}}_{{h_{n,k}}}};{{{\larrow{m}}}_{{g_{n,k}}}},{{{\rarrow{v}}}_{{h_{n,k}}}} \!+\!{{{\larrow{v}}}_{{g_{n,k}}}}\big)  \!- \!  \left| {{{\hat{h}}}_{n,k}} \right|^{2},
\end{salign}
where $ O_{h_{n,k}}$ is a normalization coefficient
\begin{salign}
O_{h_{n,k}} =  \sum\limits_{u\in \mathsf{\mathcal{U}}}   {\beta _{k}^{u}} \!\left[ \left( 1 -   {{s}_{n,u}} \right)\mathcal{CN}\big( 0;{{{\rarrow{m}}}_{{h_{n,k}}}}, {{{\rarrow{v}}}_{{h_{n,k}}}} \big)  + {{s}_{n,u}}\mathcal{CN}\big( {h_{n,k}};{{{ {m}}}_{n,k}},  {{{ {v}}}_{n,k}} \big) \right],
\end{salign}
and
\begin{align}
{v}_{n,k}= \Bigg( \frac{ 1}{{\rarrow{v}}_{h_{n,k}}} + \frac{1}{ {\larrow{v}}_{g_{n,k}} } \Bigg)^{-1},\quad
{m}_{n,k}= \Bigg( \frac{ {\rarrow{m}}_{h_{n,k}}}{{\rarrow{v}}_{h_{n,k}}} + \frac{{\larrow{m}}_{g_{n,k}}}{ {\larrow{v}}_{g_{n,k}} } \Bigg) {v}_{n,k}.
\end{align}
Then, the backward message {\small $I_{{{f}_{{h_{n,k}}}}\to {h_{n,k}}} \left( {h_{n,k}} \right)$}  can be expressed as
\begin{salign}\label{eq:EPfh2h}
I^{\scriptscriptstyle EP}_{{{f}_{{h_{n,k}}}}\to {h_{n,k}}} \left( {h_{n,k}} \right)
\notag &	= \frac{b^{\scriptscriptstyle G}\left({h_{n,k}}\right)}{ {{I}_{{h_{n,k}}  \to \! {{f}_{{h_{n,k}}}}}} \left({h_{n,k}}\right) }\\
&	\varpropto \mathcal{CN}\left( {h_{n,k}};{{{\larrow{m}}}_{{h_{n,k}}}},{{{\larrow{v}}}_{{h_{n,k}}}} \right).
\end{salign}
where
\begin{small}
\begin{align}\label{eq:MVfh2h}
{{{\larrow{v}}}_{{h_{n,k}}}}  = \Bigg( \frac{ 1}{{v}_{h_{n,k}}} - \frac{1}{ {\rarrow{v}}_{h_{n,k}} } \Bigg)^{-1},\quad
{{{\larrow{m}}}_{{h_{n,k}}}}  = \Bigg( \frac{{{{\hat{h}}}_{n,k}}}{{{v}_{{h_{n,k}}}}} - \frac{{{\rarrow{m}}_{{h_{n,k}}}}}{{{\rarrow{v}}_{{h_{n,k}}}}} \Bigg){{{\larrow{v}}}_{{h_{n,k}}}}.
\end{align}
\end{small}\normalsize
As show in Fig. \ref{FG}, the outgoing messages {\small $ \big\lbrace I^{\scriptscriptstyle EP}_{{{f}_{{h_{n,k}}}}\to {h_{n,k}}} \left( {h_{n,k}} \right), \forall n,k \big\rbrace $} are input to Part (ii).

\subsection{Message Passing in Part (ii)}
With the incoming messages  {\small $ \big\lbrace I^{\scriptscriptstyle EP}_{{{f}_{{h_{n,k}}}}\to {h_{n,k}}} \left( {h_{n,k}} \right), \forall n,k \big\rbrace $}, the message passing in Part (ii) realizes the estimation of signal $ \textbf{X} $ and noise precision $ \lambda $. The key is to deal with the observation nodes {\small $\{ f_{y_{n,l}},\forall n,l \}$}, which can be tackled with pure MF or BP-MF to achieve trade-off between complexity and performance. These two methods are elaborated in the following subsections.

\vspace{2mm}
\textbf{(1) MF based message passing } 

As the MF rule will be used to handle the observation nodes, the incoming messages to the observation nodes are the beliefs of the relevant variables. To calculate the backward message {\small $ {{I}_{f_{y_{n,l}}\to x_{\scriptscriptstyle k,l}}}( x_{\scriptscriptstyle k,l}) $},
the beliefs {\small $ \big\lbrace b^{\scriptscriptstyle G} (x_{k',l}), \forall k' \neq k \big\rbrace $}, $ b\left( \lambda  \right) $ and  {\small $ \big\lbrace b \left( {h_{n,k}} \right), \forall n,k \big\rbrace $}, which are given respectively in   (\ref{eq:blfX}), (\ref{eq:blfLmd1}) and (\ref{eq:blfHa1}), are required. So we have
\begin{salign}
 {{I}_{f_{y_{n,l}}\to x_{\scriptscriptstyle k,l}}}\left( x_{\scriptscriptstyle k,l} \right)
\notag &=\exp \left\{ \int \ln \left[ f_{y_{n,l}}\left( \lambda ,{h_{n,k}},x_{\scriptscriptstyle k,l},\forall k \right) \right] \prod\limits_{{k}'\ne k}{{{b}}^{\scriptscriptstyle G}\left( x_{k',l} \right)} \right. \\
\notag & \left.\qquad \qquad \cdot b\left( \lambda  \right)\prod\limits_{n=1}^{N}{b\left( {h_{n,k}} \right)}  d\lambda d\mathbf{h}{d{{\mathbf{x}}}}/{x_{\scriptscriptstyle k,l}} \right\} \\
& \varpropto \mathcal{CN}\left( x_{\scriptscriptstyle k,l}; {{{\overset{\scriptscriptstyle\twoheadleftarrow}{m}}}_{{{x}_{\scriptscriptstyle k,l}}}},{{{\overset{\scriptscriptstyle\twoheadleftarrow}{v}}}_{{{x}_{\scriptscriptstyle k,l}}}} \right),
\end{salign}
where
\begin{salign}\label{eq:MVfy2x}
{{\overset{\scriptscriptstyle\twoheadleftarrow}{m}}_{{{x}_{\scriptscriptstyle k,l}}}}=\frac{ {{{\hat{h}}}_{n,k}}^{H}\Big( y_{n,l}-\sum\limits_{{k}'\ne k}{{{{\hat{h}}}_{n,k'}}\hat{x}_{k',l}} \Big)}{{{\left| {{{\hat{h}}}_{n,k}} \right|}^{2}}+{{v}_{{h_{n,k}}}}},\
{{\overset{\scriptscriptstyle\twoheadleftarrow}{v}}_{{{x}_{\scriptscriptstyle k,l}}}} = \frac{1}{{\hat\lambda }\Big( {{ \left| {{{\hat{h}}}_{n,k}} \right|}^{2}}+{{v}_{{h_{n,k}}}} \Big)}.
\end{salign}
Then, the belief  of $ x_{\scriptscriptstyle k,l}$ can be updated by
\begin{salign}
\notag {{b}}\left( x_{\scriptscriptstyle k,l} \right) & \varpropto  f_{x_{\scriptscriptstyle k,l}}\left( x_{\scriptscriptstyle k,l} \right)\prod\limits_{n=1}^{N}{{{I}_{f_{y_{n,l}}\to x_{\scriptscriptstyle k,l}}}\left( x_{\scriptscriptstyle k,l} \right)}  \\
& \triangleq \sum\limits_{q\in \mathsf{\mathcal{X}}}{\beta _{\scriptscriptstyle k,l}^{q}\delta \left( x_{\scriptscriptstyle k,l}-q \right)} ,
\end{salign}
where
\begin{salign}\label{eq:Beta_klq}
\beta _{\scriptscriptstyle k,l}^{q} = \frac{\mathcal{CN}\left( q; {{{\larrow{m}}}_{{{x}_{\scriptscriptstyle k,l}}}},{{{\larrow{v}}}_{{{x}_{\scriptscriptstyle k,l}}}} \right)}{\sum\limits_{q'\in \mathsf{\mathcal{X}}}{\mathcal{CN}\left( q'; {{\larrow{m}}_{{{x}_{\scriptscriptstyle k,l}}}},{{\larrow{v}}_{{{x}_{\scriptscriptstyle k,l}}}} \right)}} ,
\end{salign}
with
\begin{salign}\label{eq:SumMVx}
{{\larrow{v}}_{{{x}_{\scriptscriptstyle k,l}}}} = {{\Bigg( \sum\limits_{n=1}^{N}{ \frac{1} {{{\overset{\scriptscriptstyle\twoheadleftarrow}{v}}}_{{{x}_{\scriptscriptstyle k,l}}}} } \Bigg)}^{-1}},\quad
{{\larrow{m}}_{{{x}_{\scriptscriptstyle k,l}}}}=\Bigg( \sum\limits_{n=1}^{N}{\frac{{{{\overset{\scriptscriptstyle\twoheadleftarrow}{m}}}_{{{x}_{\scriptscriptstyle k,l}}}}}{{{{\overset{\scriptscriptstyle\twoheadleftarrow}{v}}}_{{{x}_{\scriptscriptstyle k,l}}}}}} \Bigg){{\larrow{v}}_{{{x}_{\scriptscriptstyle k,l}}}}.
\end{salign}
Note that {\small $\left\lbrace b \left( x_{\scriptscriptstyle k,l} \right)  ,\forall k,l \right\rbrace $} in the last iteration are used for soft demodulation and decoding. As $ b \left( x_{\scriptscriptstyle k,l} \right) $ is no longer Gaussian, it can be approximated to be Gaussian using moment matching, i.e.,
\begin{salign}\label{eq:blfX}
\notag b \left( x_{\scriptscriptstyle k,l} \right)  & \approx b^{\scriptscriptstyle G} \left( x_{\scriptscriptstyle k,l} \right) \\
\notag & = \text{Proj}_{G}\left\{ f_{x_{\scriptscriptstyle k,l}}\left( x_{\scriptscriptstyle k,l} \right)\prod\limits_{n=1}^{N}{{{I}_{f_{y_{n,l}}\to x_{\scriptscriptstyle k,l}}}\left( x_{\scriptscriptstyle k,l} \right)}  \right\} \\
& \triangleq \mathcal{CN}\left( x_{\scriptscriptstyle k,l};\hat x_{\scriptscriptstyle k,l},{{v}_{{{x}_{\scriptscriptstyle k,l}}}} \right) ,
\end{salign}
where
\begin{salign}
\label{eq:Mx} \hat x_{\scriptscriptstyle k,l}  &= {{\big\langle {x}_{\scriptscriptstyle k,l}  \big\rangle }_{{{b}}({x}_{\scriptscriptstyle k,l})}}  =\sum\limits_{q\in \mathsf{\mathcal{X}}} q \beta _{\scriptscriptstyle k,l}^{q} ,\\
\label{eq:Vx}  {{v}_{{{x}_{\scriptscriptstyle k,l}}}} &= {{\text{Var}\big[ {x}_{\scriptscriptstyle k,l}  \big] }_{{{b}}({x}_{\scriptscriptstyle k,l})}} = \sum\limits_{q\in \mathsf{\mathcal{X}}}{ \left| q \right|^{2} } \beta _{\scriptscriptstyle k,l}^{q} - {{\left| \hat x_{\scriptscriptstyle k,l} \right|}^{2}},
\end{salign}

With  the updated{\small $ \big\lbrace b^{\scriptscriptstyle G}(x_{k,l}), \forall k,l \big\rbrace $} and {\small $ \big\lbrace b \left( {h_{n,k}} \right), \forall n,k \big\rbrace $} in (\ref{eq:blfHa1}), the forward message $ {{I}_{f_{y_{n,l}}\to \lambda}}( \lambda) $ can be calculated by
\begin{salign}
 {{I}_{f_{y_{n,l}}\to \lambda }}\left( \lambda  \right)
\notag & =\exp \left\{ \int\ln \left[ f_{y_{n,l}}\left( \lambda ,{h_{n,k}},x_{\scriptscriptstyle k,l},\forall k \right) \right]\prod\limits_{n=1}^{N}{{{b}}\left( {h_{n,k}} \right)}\right. \\
\notag & \left.\qquad \qquad \cdot \prod\limits_{k=1}^{K}{{{b}}^{\scriptscriptstyle G}\left( x_{\scriptscriptstyle k,l} \right)}d\mathbf{h}d{{\mathbf{x}}} \right\} \\
& \varpropto\lambda \exp \left\{ -\lambda A_{y_{n,l}} \right\} ,
\end{salign}
where
\begin{salign}
A_{y_{n,l}}=\sum\limits_{k=1}^{K}{\left[ {{\left| {{{\hat{h}}}_{n,k}} \right|}^{2}}{{v}_{{{x}_{\scriptscriptstyle k,l}}}}+{{\left| \hat x_{\scriptscriptstyle k,l} \right|}^{2}}{{v}_{{h_{n,k}}}}+{{v}_{{{x}_{\scriptscriptstyle k,l}}}}{{v}_{{h_{n,k}}}} \right]} + {\Big| y_{n,l}-\sum\limits_{k=1}^{K}{{{{\hat{h}}}_{n,k}}\hat x_{\scriptscriptstyle k,l}} \Big|^{2}} .
\end{salign}
Then, the belief of $ \lambda $ can be updated by
\begin{salign}\label{eq:blfLmd1}
\notag  b\left( \lambda  \right) & \varpropto \prod\limits_{l=1}^{L}{\prod\limits_{n=1}^{N}{{{I}_{f_{y_{n,l}}\to \lambda }}\left( \lambda  \right)}}{{f}_{\lambda }}\left( \lambda  \right) \\
& \propto {{\lambda}^{NL-1}}\exp \left\{ -\lambda \sum\limits_{l=1}^{L}{\sum\limits_{n=1}^{N}{A_{y_{n,l}}}} \right\}.
\end{salign}
Thus, the noise  precision can be calculated as
\begin{equation} \label{eq:lambda1}
\hat{\lambda} ={{\big\langle \lambda  \big\rangle }_{b\left( \lambda  \right)}}
 = \frac{NL}{\sum\limits_{L=1}^{L}{\sum\limits_{n=1}^{N}{A_{y_{n,l}}}}}.
\end{equation}

\begin{algorithm} [t]
	\caption{ MF based  MUD joint with UAD and CE}
	\label{alg:MF-MP}
	{\bf Input:}
	{\small $ \textbf{Y},  p\left( \lambda  \right),   \left\lbrace p\left( {z}_{\scriptscriptstyle k}\right) \right\rbrace,   \left\lbrace p\left( {g_{n,k}}\right) \right\rbrace,   \left\lbrace p\left(  x_{\scriptscriptstyle k,l} \right) \right\rbrace.$ } \\
	\textbf{Initialize:} {\small $ \hat{\lambda}; \big\lbrace {{\rarrow{m}}}_{{h_{n,k}}}, {{\rarrow{v}}}_{{h_{n,k}}} \big\rbrace  ; \  \forall k,l, \hat x_{\scriptscriptstyle k,l}=q, q \in \mathcal{X}$}\\  	
	1: \textbf{for} {\small $ i=1:N_{O_{itr}} $} (Outer iteration) \\	
	2: \hspace*{0.15 in} $ \forall k,u $:  update {\small $ \beta _{k}^{u} $} by (\ref{eq:BetaKU}). \\	
	3: \hspace*{0.15 in} $ \forall n,k $:  update {\small $ {{{\hat{h}}}_{n,k}} $} and {\small $ v_{{h_{n,k}}} $} by (\ref{eq:epMh}) and (\ref{eq:epVh}). \\	
	4: \hspace*{0.15 in} $ \forall n,k $:  update {\small $ {{{\larrow{m}}}_{{h_{n,k}}}} $} and {\small $ {{{\larrow{v}}}_{{h_{n,k}}}} $} by (\ref{eq:MVfh2h}). \\
	5: \hspace*{0.15 in} \textbf{for} {\small $ j=1:N_{I_{itr}} $} (Inner iteration) \\ 	
	6: \hspace*{0.3 in} $ \forall k,l $:  update {\small $ {{\overset{\scriptscriptstyle\twoheadleftarrow}{m}}_{{{x}_{\scriptscriptstyle k,l}}}} $} and {\small $ {{\overset{\scriptscriptstyle\twoheadleftarrow}{v}}_{{{x}_{\scriptscriptstyle k,l}}}} $} by (\ref{eq:MVfy2x}). 	\\
	7: \hspace*{0.3 in} $ \forall k,l,q $:  update {\small $ \beta _{\scriptscriptstyle k,l}^{q} $} by (\ref{eq:Beta_klq}).\\ 	
	8: \hspace*{0.3 in} $ \forall k,l $:  update {\small $ \hat x_{\scriptscriptstyle k,l} $} and {\small $ {{v}_{{{x}_{\scriptscriptstyle k,l}}}} $} by (\ref{eq:Mx}) and (\ref{eq:Vx}). \\	
	9: \hspace*{0.3 in}   update  {\small $ \hat{\lambda} $} by (\ref{eq:lambda1}). \\	
	10: \hspace*{0.25 in} $ \forall n,k $:  update {\small $ {{\overset{\scriptscriptstyle\twoheadrightarrow}{m}}_{{h_{n,k}}}} $} and {\small $ {{\overset{\scriptscriptstyle\twoheadrightarrow}{v}}_{{h_{n,k}}}} $} by (\ref{eq:MVfy2h}). \\	
	11: \hspace*{0.25 in} $ \forall n,k $:  update {\small $ {{{\rarrow{m}}}_{{h_{n,k}}}} $} and {\small $  {{{\rarrow{v}}}_{{h_{n,k}}}}$} by (\ref{eq:MVh2fh}). \\	
	12: \hspace*{0.25 in} $ \forall n,k $:  update {\small $ {{{\hat{h}}}_{n,k}} $} and {\small $ v_{{h_{n,k}}} $} by (\ref{eq:MVh}).\\	
	13: \hspace*{0.15 in} \textbf{end} \\ 	
	14: \textbf{end}\\ 	
	{\bf Output:} Posterior distributions:
	{\small $  b\left(z_{k}\right)= \sum\limits_{u\in \mathsf{\mathcal{U}}}{\beta _{k}^{u}\delta \left( {{z}_{\scriptscriptstyle k}}-u \right)},\forall k;$}\\
	\hspace*{0.55 in}{\small $ \hat h_{n,k}, \forall n,k; \qquad  {{b}}\left( x_{\scriptscriptstyle k,l} \right) = \sum\limits_{q\in \mathsf{\mathcal{X}}}{\beta _{\scriptscriptstyle k,l}^{q}\delta \left( x_{\scriptscriptstyle k,l}-q \right)}, \forall k,l.$}
\end{algorithm}

With the belief {\small $\left\lbrace b\left({h}_{n',k}\right), \forall n' \neq n \right\rbrace $}, the updated belief {\small $ \left\lbrace b^{\scriptscriptstyle G}\left( x_{\scriptscriptstyle k,l} \right), \forall k,l \right\rbrace $}  and $ b ( \lambda ) $,  the forward message {\small $ {{I}_{f_{y_{n,l}}\to {h_{n,k}}}}\left( {h_{n,k}} \right) $} can be calculated by
\begin{salign}
\notag  {{I}_{f_{y_{n,l}}\to {h_{n,k}}}}\left( {h_{n,k}} \right)
&=\exp \left\{ \!\! \int \ln \left[ f_{y_{n,l}}\left( \lambda ,{h_{n,k}},x_{\scriptscriptstyle k,l},\forall k \right) \right]b\left( \lambda  \right)\prod\limits_{{n}'\ne n}{b\left( {{h}_{n',k}} \right)} \right. \\
\notag & \left.\qquad \qquad \cdot \prod\limits_{k=1}^{K}{{{b}}^{\scriptscriptstyle G}\left( x_{k,l} \right)}d\lambda d{{\mathbf{x}}}{d\mathbf{h}}/{{h_{n,k}}} \right\} \\
& \varpropto \mathcal{CN}\left( h_{n,k}; {{\overset{\scriptscriptstyle\twoheadrightarrow}{m}}_{{h_{n,k}}}} ,{{\overset{\scriptscriptstyle\twoheadrightarrow}{v}}_{{h_{n,k}}}} \right),
\end{salign}
where
\begin{salign}\label{eq:MVfy2h}
{{\overset{\scriptscriptstyle\twoheadrightarrow}{m}}_{{h_{n,k}}}} = \frac{\hat x_{\scriptscriptstyle k,l}^{H}  \Big( y_{n,l}-\sum\limits_{{n}'\ne n}{{{{\hat{h}}}_{n',k}}\hat x_{\scriptscriptstyle k,l}} \Big)}{ {{\left| \hat x_{\scriptscriptstyle k,l} \right|}^{2}}+{{v}_{{{x}_{\scriptscriptstyle k,l}}}} }, \quad
{{\overset{\scriptscriptstyle\twoheadrightarrow}{v}}_{{h_{n,k}}}} = \frac{1}{\hat{\lambda }\left( {{\left| \hat x_{\scriptscriptstyle k,l} \right|}^{2}}+{{v}_{{{x}_{\scriptscriptstyle k,l}}}} \right)}.
\end{salign}
Thus, the forward message $ {{I}_{{h_{n,k}}\to {{f}_{{h_{n,k}}}}}}\left( {h_{n,k}} \right) $, which is used in Part (i), is given by
\begin{salign}\label{eq:htofh1}
\notag {{I}_{{h_{n,k}}\to {{f}_{{h_{n,k}}}}}}\left( {h_{n,k}} \right) & =\prod\limits_{n=1}^{N}{ {{I}_{f_{y_{n,l}}\to {h_{n,k}}}}\left( {h_{n,k}} \right) } \\
& \varpropto \mathcal{CN}\left( {h_{n,k}};{{{\rarrow{m}}}_{{h_{n,k}}}},{{{\rarrow{v}}}_{{h_{n,k}}}} \right),
\end{salign}
where
\begin{salign}\label{eq:MVh2fh}
{{{\rarrow{v}}}_{{h_{n,k}}}} ={{\left( \sum\limits_{n=1}^{N}{{{{\overset{\scriptscriptstyle\twoheadrightarrow}{v}}_{{h_{n,k}}}}}^{-1}} \right)}^{-1}}, \quad
{{{\rarrow{m}}}_{{h_{n,k}}}}  = \Bigg( \sum\limits_{n=1}^{N}{\frac{{{\overset{\scriptscriptstyle\twoheadrightarrow}{m}}_{{h_{n,k}}}}}{{{\overset{\scriptscriptstyle\twoheadrightarrow}{v}}_{{h_{n,k}}}}}} \Bigg){{{\rarrow{v}}}_{{h_{n,k}}}}.
\end{salign}
Then, with {\small $ I^{\scriptscriptstyle EP}_{{{f}_{{h_{n,k}}}}\to {h_{n,k}}} \big( {h_{n,k}} \big) \varpropto \mathcal{CN}\big( {h_{n,k}};{{{\larrow{m}}}_{{h_{n,k}}}},{{{\larrow{v}}}_{{h_{n,k}}}} \big) $}  from Part (i) in (\ref{eq:EPfh2h}) and the updated message {\small $ {{I}_{{h_{n,k}}\to {{f}_{{h_{n,k}}}}}}\left( {h_{n,k}} \right) $} , the belief of $ h_{n,k} $ is updated by
\begin{salign}\label{eq:blfHa1}
\notag b\left({h_{n,k}}\right) & \varpropto I^{\scriptscriptstyle EP}_{{{f}_{{h_{n,k}}}}\! \to  {h_{n,k}}} \left({h_{n,k}}\right){{I}_{{h_{n,k}}  \to \! {{f}_{{h_{n,k}}}}}} \left({h_{n,k}}\right) \\
& \varpropto \mathcal{CN}\left( {h_{n,k}}; {{{\hat{h}}}_{n,k}},v_{{h_{n,k}}} \right),
\end{salign}
where
\begin{salign}\label{eq:MVh}
v_{{h_{n,k}}}  = \Bigg( \frac{1}{{\rarrow{v}}_{h_{n,k}}} + \frac{1}{{\larrow{v}}_{h_{n,k}}} \Bigg)^{-1}, \quad
{{{\hat{h}}}_{n,k}}  = \Bigg( \frac{{\rarrow{m}}_{h_{n,k}}}{{\rarrow{v}}_{h_{n,k}}} + \frac{{\larrow{m}}_{h_{n,k}}}{{\larrow{v}}_{h_{n,k}}} \Bigg) v_{{h_{n,k}}}.
\end{salign}

The algorithm of message passing  based MUD joint with UAD and CE, described in Section III.B and  Section III.C.(1), is summarized in  {Algorithm \ref{alg:MF-MP}}. \vspace{2mm}

\textbf{(2) BP and MF based message passing} 

\begin{figure}
	\centering
	\includegraphics[width=3.2in]{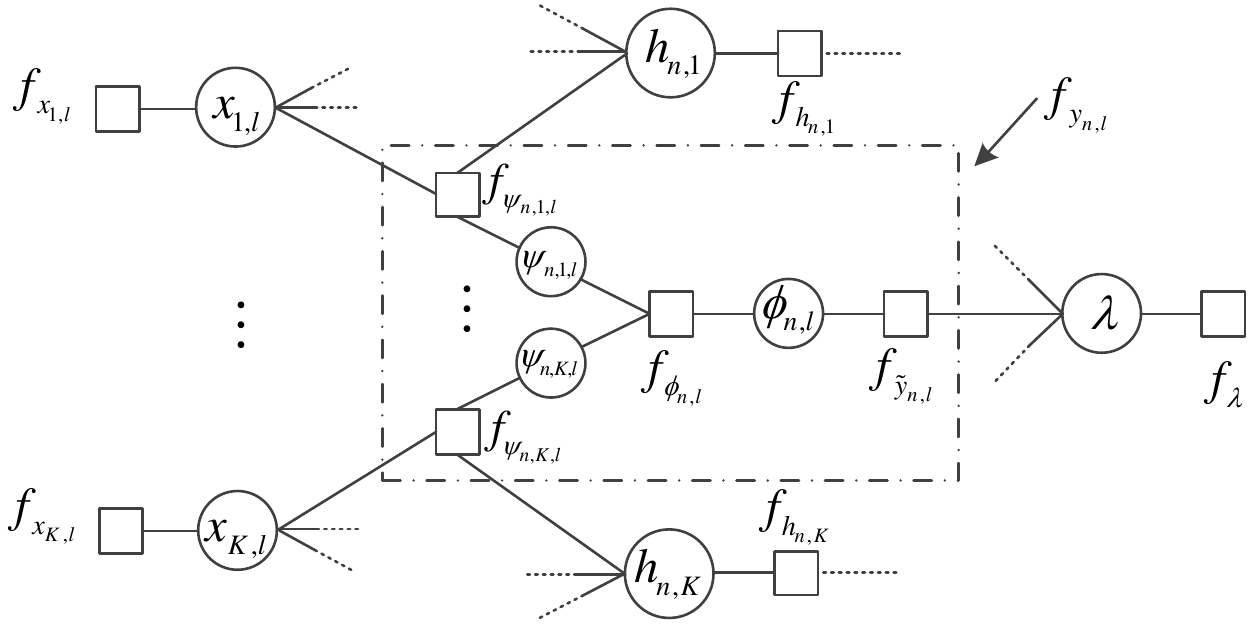}
	\caption{Factor graph representation of (\ref{eq:factorfy}). }\label{FGfy}
\end{figure}

We note that the observation factors $\{ f_{y_{n,l}},\forall n,l \}$ are functions of a number of variables in the form of multiplication and multi-signal summation, where MF is not effective to deal with. To overcome the drawback of MF, we decompose  each observation factor into some sub-factors, which enables the use of both BP and MF to tackle different sub-factors, leading to considerable performance improvement. Let's define {\small $ \psi _{n,k,l} = {h_{n,k}}x_{\scriptscriptstyle k,l} $} and  {\small $ \phi _{n,l} = \sum\nolimits_{k=1}^{K}{\psi _{n,k,l}} $}, and these hard constrains  can be represented by factors {\small $ f_{\psi _{\scriptscriptstyle n,k,l}}( \psi _{n,k,l},{h_{n,k}},x_{\scriptscriptstyle k,l}) =  \delta (\psi _{n,k,l}-{h_{n,k}}x_{\scriptscriptstyle k,l}) $} and {\small $ f_{\phi _{\scriptscriptstyle n,l}}(\phi _{n,l},\psi _{n,k,l}, \forall k) = \delta(\phi _{n,l}-\sum\nolimits_{k=1}^{K}{\psi _{n,k,l}}) $}, respectively.  In addition, define {\small $ f_{\tilde{y} _{\scriptscriptstyle n,l}} (\phi _{n,l} ,\lambda ) = \mathcal{CN}( y_{n,l};\phi _{n,l}, {{\lambda }^{-1}})$}. Then the original observation factor {\small $ f_{y_{n,l}} ( \lambda ,{h_{n,k}},x_{\scriptscriptstyle k,l},\forall k )$}  can be expressed as
\begin{salign}
f_{y_{n,l}}\left( \lambda ,{h_{n,k}},x_{\scriptscriptstyle k,l},\forall k \right)
 = \!\!\! \iint \!\!\! f \left( \lambda , \phi _{\scriptscriptstyle n,l},{\psi _{\scriptscriptstyle n,k,l}}, {h_{n,k}},x_{\scriptscriptstyle k,l},\forall k \right)     d\phi _{n,l}d\pmb{\psi }_{n,l}.
\end{salign}
where
\begin{salign}\label{eq:factorfy}
	f \left( \lambda ,\phi _{\scriptscriptstyle n,l}, {\psi _{\scriptscriptstyle n,k,l}}, {h_{n,k}},x_{\scriptscriptstyle k,l},\forall k \right)
	\notag &  =  f_{\tilde{y} _{\scriptscriptstyle n,l}}\left(\phi _{n,l} ,\lambda \right) f_{\phi _{\scriptscriptstyle n,l}} \left(\phi _{n,l},\psi _{n,k,l} \, ,\forall k \right)\\
	& \qquad \cdot \prod\limits_{k=1}^{K}  \!{f_{\psi _{\scriptscriptstyle n,k,l}}\left(  \psi _{n,k,l},{h_{n,k}},x_{\scriptscriptstyle k,l} \right) }     .
\end{salign}
The  factor graph representation of (\ref{eq:factorfy}) is shown in the dashed box in Fig. \ref{FGfy},  where MF is performed at the function node {\small $ f_{\tilde{y} _{\scriptscriptstyle n,l}}$} for noise precision estimation, and BP is performed at the function nodes {\small $ f_{\phi _{\scriptscriptstyle n,l}} $} and {\small $\big\lbrace  f_{\psi _{\scriptscriptstyle n,k,l}} , \forall k\big\rbrace $}  for multi-signal detection. Next, we detail the message computations in the factor graph shown in Fig. \ref{FGfy}.

With {\small $ {{I}_{f_{\psi _{\scriptscriptstyle n,k,l}}\to {h_{n,k}}}} \big( {h_{n,k}} \big) \varpropto \mathcal{CN} \big( {h_{n,k}};  {{\overset{\scriptscriptstyle\twoheadrightarrow}{m}}_{{h_{n,k}}}},{{\overset{\scriptscriptstyle\twoheadrightarrow}{v}}_{{h_{n,k}}}} \big)$} given in (\ref{eq:fpsi2h}) and the belief {\small $  b\left({h_{n,k}}\right)$}, the backward message {\small $ {{I}_{{h_{n,k}}\to f_{\psi _{\scriptscriptstyle n,k,l}}}}\big( {h_{n,k}} \big) $} can be expressed as
\begin{salign}\label{eq:h2fpsi}
  {{I}_{{h_{n,k}}\to f_{\psi _{\scriptscriptstyle n,k,l}}}}\left( {h_{n,k}} \right)   =\frac{b\left( {h_{n,k}} \right)}{{{I}_{f_{\psi _{\scriptscriptstyle n,k,l}}\to {h_{n,k}}}}\left( {h_{n,k}} \right)}
 \varpropto \mathcal{CN}\left( {h_{n,k}};{{\overset{\scriptscriptstyle\twoheadleftarrow}{m}}_{{h_{n,k}}}},{{\overset{\scriptscriptstyle\twoheadleftarrow}{v}}_{{h_{n,k}}}} \right),
\end{salign}
where
\begin{salign}\label{eq:MVh2fpsi}
{{\overset{\scriptscriptstyle\twoheadleftarrow}{v}}_{{h_{n,k}}}}=
\Bigg( \frac{1}{{{v}_{{h_{n,k}}}}} - \frac{1}{{{\overset{\scriptscriptstyle\twoheadrightarrow}{v}}_{{h_{n,k}}}}}  \Bigg)^{-1}, \quad
{{\overset{\scriptscriptstyle\twoheadleftarrow}{m}}_{{h_{n,k}}}}=\Bigg( \frac{{{{\hat{h}}}_{n,k}}}{{{v}_{{h_{n,k}}}}} - \frac{{{\overset{\scriptscriptstyle\twoheadrightarrow}{m}}_{{h_{n,k}}}}}{{{\overset{\scriptscriptstyle\twoheadrightarrow}{v}}_{{h_{n,k}}}}}  \Bigg) {{\overset{\scriptscriptstyle\twoheadleftarrow}{v}}_{{h_{n,k}}}}.
\end{salign}
Then, with {\small $ {{I}_{\psi _{n,k,l}\to f_{\psi _{\scriptscriptstyle n,k,l}}}} \!\! \big(\psi _{n,k,l} \big) \varpropto \mathcal{CN}\big( \psi _{n,k,l};{{{\larrow{m}}}_{{{\psi }_{n,k,l}}}},{{{\larrow{v}}}_{{{\psi }_{n,k,l}}}} \big) $ } given in (\ref{eq:psi2fpsi}), the backward message $ {{I}_{f_{\psi _{\scriptscriptstyle n,k,l}}\to x_{\scriptscriptstyle k,l}}}( x_{\scriptscriptstyle k,l}) $ is given by
\begin{salign}
\notag &{{I}_{f_{\psi _{\scriptscriptstyle n,k,l}}\to x_{\scriptscriptstyle k,l}}}\left( x_{\scriptscriptstyle k,l} \right) \\
\notag & = {\Big\langle   f_{\psi _{\scriptscriptstyle n,k,l}} \big(\psi _{n,k,l},{h_{n,k}},x_{\scriptscriptstyle k,l}\big) \Big\rangle}_{{{I}_{{h_{n,k}}\to f_{\psi _{\scriptscriptstyle n,k,l}}}} \big({h_{n,k}}\big)   {{I}_{\psi _{n,k,l}\to f_{\psi _{\scriptscriptstyle n,k,l}}}} \big(\psi _{n,k,l} \big)} \\
& = \mathcal{CN}\Big( {{{\larrow{m}}}_{{{\psi }_{n,k,l}}}};x_{\scriptscriptstyle k,l}{{\overset{\scriptscriptstyle\twoheadleftarrow}{m}}_{{h_{n,k}}}},{{{\larrow{v}}}_{{{\psi }_{n,k,l}}}}+{{\big| x_{\scriptscriptstyle k,l} \big|}^{2}}{{\overset{\scriptscriptstyle\twoheadleftarrow}{m}}_{{h_{n,k}}}} \Big).
\end{salign}
Thus, the belief of $ x_{\scriptscriptstyle k,l} $ is updated by
\begin{salign}
\notag b\left( x_{\scriptscriptstyle k,l} \right)& \varpropto f_{x_{\scriptscriptstyle k,l}}\left( x_{\scriptscriptstyle k,l} \right)\prod\limits_{n=1}^{N} {I_{f_{\psi _{\scriptscriptstyle n,k,l}}\to x_{\scriptscriptstyle k,l}}\left( x_{\scriptscriptstyle k,l} \right)} \\
& \propto \sum\limits_{q\in \mathsf{\mathcal{X}}}{\beta _{\scriptscriptstyle k,l}^{q}\delta \left( x_{\scriptscriptstyle k,l}-q \right)} ,
\end{salign}
where
\begin{salign}
\beta _{\scriptscriptstyle k,l}^{q} = \!\! \frac{\prod\limits_{n=1}^{N} \mathcal{CN}\left( {{{\larrow{m}}}_{{{\psi }_{n,k,l}}}};q{{\overset{\scriptscriptstyle\twoheadleftarrow}{m}}_{{h_{n,k}}}},{{{\larrow{v}}}_{{{\psi }_{n,k,l}}}}+{{\left| q \right|}^{2}}{{\overset{\scriptscriptstyle\twoheadleftarrow}{m}}_{{h_{n,k}}}} \right)} {\sum\limits_{q'\in \mathsf{\mathcal{X}}}  \prod\limits_{n=1}^{N}  \mathcal{CN}\left( {{{\larrow{m}}}_{{{\psi }_{n,k,l}}}};q'{{\overset{\scriptscriptstyle\twoheadleftarrow}{m}}_{{h_{n,k}}}},{{{\larrow{v}}}_{{{\psi }_{n,k,l}}}}+{{\left| q' \right|}^{2}}{{\overset{\scriptscriptstyle\twoheadleftarrow}{m}}_{{h_{n,k}}}} \right)}.
\end{salign}
Note that {\small $\left\lbrace b \left( x_{\scriptscriptstyle k,l} \right)  ,\forall k,l \right\rbrace $} in the last iteration are used for soft demodulation and decoding.

The forward message {\small $ {{I}_{x_{\scriptscriptstyle k,l}\to f_{\psi _{\scriptscriptstyle n,k,l}}}} \big( x_{\scriptscriptstyle k,l} \big) $} can be expressed as
\begin{salign}\label{eq:x2fpsi}
{{I}_{x_{\scriptscriptstyle k,l}\to f_{\psi _{\scriptscriptstyle n,k,l}}}} \big( x_{\scriptscriptstyle k,l} \big)
\notag& = f_{x_{\scriptscriptstyle k,l}}\big( x_{\scriptscriptstyle k,l} \big)\prod\limits_{n' \neq n} {I_{f_{\psi _{\scriptscriptstyle n'k,l}}\to x_{\scriptscriptstyle k,l}}\big( x_{\scriptscriptstyle k,l} \big)} \\
& = \sum\limits_{q\in \mathsf{\mathcal{X}}}{\gamma^{q}_{\scriptscriptstyle k,l}\delta \left( x_{\scriptscriptstyle k,l}-q \right)},
\end{salign}
where
\begin{salign}\label{eq:Gammaqk,l}
\gamma^{q}_{\scriptscriptstyle k,l} =  \frac{1}{Q} \prod\limits_{n' \neq n} \mathcal{CN}\left( {{{\larrow{m}}}_{{{\psi }_{n',k,l}}}};q{{\overset{\scriptscriptstyle\twoheadleftarrow}{m}}_{{{h}_{n',k}}}},{{{\larrow{v}}}_{{{\psi }_{n',k,l}}}}+{{\left| q \right|}^{2}}{{\overset{\scriptscriptstyle\twoheadleftarrow}{m}}_{{{h}_{n',k}}}} \right).
\end{salign}
Then, the forward message {\small $ {{I}_{f_{\psi _{\scriptscriptstyle n,k,l}}\to \psi _{n,k,l}}} \big(\psi _{n,k,l}\big) $}  is given by
\begin{salign}
\notag &{{I}_{f_{\psi _{\scriptscriptstyle n,k,l}}\to \psi _{n,k,l}}}\big( \psi _{n,k,l} \big)\\
\notag & = {\Big\langle  f_{\psi _{\scriptscriptstyle n,k,l}} \big(\psi _{n,k,l},{h_{n,k}},x_{\scriptscriptstyle k,l}\big) \Big\rangle}_{{{I}_{x_{\scriptscriptstyle k,l}\to f_{\psi _{\scriptscriptstyle n,k,l}}}} \big( x_{\scriptscriptstyle k,l}\big)  {{I}_{{h_{n,k}}\to f_{\psi _{\scriptscriptstyle n,k,l}}}} \big( {h_{n,k}}\big)} \\
& = \sum\limits_{q\in \mathsf{\mathcal{X}}}{\gamma^{q}_{\scriptscriptstyle k,l}}{{\left| q \right|}^{2}}\mathcal{CN}\left( \psi _{n,k,l};q{{\overset{\scriptscriptstyle\twoheadleftarrow}{m}}_{{h_{n,k}}}},{{\left| q \right|}^{2}}{{\overset{\scriptscriptstyle\twoheadleftarrow}{v}}_{{h_{n,k}}}} \right) ,
\end{salign}
which is not Gaussian. To reduce the complexity, it is approximated to be Gaussian by  moment matching, i.e.
\begin{salign}\label{eq:fpsi2psi}
\notag  I_{f_{\psi _{\scriptscriptstyle n,k,l}}\to \psi _{n,k,l}}\left( \psi _{n,k,l} \right)
&\approx\text{Proj}_{G}\left\{ {{I}_{f_{\psi _{\scriptscriptstyle n,k,l}}\to \psi _{n,k,l}}}\left( \psi _{n,k,l} \right) \right\} \\
& \triangleq \mathcal{CN}\left( \psi _{n,k,l};{{{\rarrow{m}}}_{{{\psi }_{n,k,l}}}},{{{\rarrow{v}}}_{{{\psi }_{n,k,l}}}} \right),
\end{salign}
where
\begin{salign}\label{eq:Mfpsi2psiA2}
\notag  {{{\rarrow{m}}}_{{{\psi }_{n,k,l}}}}
&={{\big \langle \psi _{n,k,l} \big \rangle }_{{{I}_{f_{\psi _{\scriptscriptstyle n,k,l}}\to \psi _{n,k,l}}}\left( \psi _{n,k,l} \right)}}\\
&=\frac{\sum\limits_{q\in \mathsf{\mathcal{X}}}{\gamma^{q}_{\scriptscriptstyle k,l}}{{\left| q \right|}^{2}}\cdot {{\overset{\scriptscriptstyle\twoheadleftarrow}{m}}_{{h_{n,k}}}} q }{\sum\limits_{q\in \mathsf{\mathcal{X}}}{\gamma^{q}_{\scriptscriptstyle k,l}}{{\left| q \right|}^{2}}} ,
\end{salign}
\begin{salign}\label{eq:Vfpsi2psiA2}
{{{\rarrow{v}}}_{{{\psi }_{n,k,l}}}}
\notag & ={{ \text{Var}\big[ \psi _{n,k,l} \big]}_{{{I}_{f_{\psi _{\scriptscriptstyle n,k,l}}\to \psi _{n,k,l}}}\left( \psi _{n,k,l} \right)}} \\
& =\frac{\sum\limits_{q\in \mathsf{\mathcal{X}}}{\gamma^{q}_{\scriptscriptstyle k,l}}{{\left| q \right|}^{4}}\Big( {{\big| {{\overset{\scriptscriptstyle\twoheadleftarrow}{m}}_{{h_{n,k}}}}\big|}^{2}}+{{\overset{\scriptscriptstyle\twoheadleftarrow}{v}}_{{h_{n,k}}}} \Big)}{\sum\limits_{q\in \mathsf{\mathcal{X}}}{\gamma^{q}_{\scriptscriptstyle k,l}}{{\left| q \right|}^{2}}}-{{\left| {{{\rarrow{m}}}_{{{\psi }_{n,k,l}}}} \right|}^{2}}.
\end{salign}
Thus,  the forward message {\small $ {{I}_{f_{\phi _{\scriptscriptstyle n,l}}\to \phi _{n,l}}} \big( \phi _{n,l}\big) $} is given by
\begin{salign}\label{eq:fphi2phi}
{{I}_{f_{\phi _{\scriptscriptstyle n,l}}\to \phi _{n,l}}}\left( \phi _{n,l} \right)
\notag &= {\Big\langle   f_{\phi _{\scriptscriptstyle n,l}} \big( \phi _{n,l},\psi _{n,k,l},\forall k \big) \Big\rangle}_{\prod\limits_{k=1}^{K}{I_{f_{\psi _{\scriptscriptstyle n,k,l}}\to \psi _{n,k,l}} \big( \psi _{n,k,l} \big)}}\\
\notag & =\mathcal{CN}\Big( \phi _{n,l};\sum_{k=1}^{K}{{{{\rarrow{m}}}_{{{\psi }_{n,k,l}}}}},\sum_{k=1}^{K}{{{{\rarrow{v}}}_{{{\psi }_{n,k,l}}}}} \Big) \\
& \triangleq \mathcal{CN}\left( \phi _{n,l};{{{\rarrow{m}}}_{{{\phi }_{n,l}}}},{{{\rarrow{v}}}_{{{\phi }_{n,l}}}} \right) .
\end{salign}

With {\small $ {{I}_{f_{\tilde{y} _{\scriptscriptstyle n,l}}\to \phi _{n,l}}} \big(\phi _{n,l}\big) $}   in (\ref{eq:fy2phi}), the belief of $ \phi _{n,l} $ can be updated by
\begin{salign}\label{blfphi}
\notag b\left( \phi _{n,l} \right)& \varpropto {{I}_{f_{\phi _{\scriptscriptstyle n,l}}\to \phi _{n,l}}}\left( \phi _{n,l} \right){{I}_{f_{\tilde{y} _{\scriptscriptstyle n,l}}\to \phi _{n,l}}}\left( \phi _{n,l} \right) \\
\notag & \varpropto \mathcal{CN}\left( \phi _{n,l};\frac{{{{\rarrow{m}}}_{{{\phi }_{n,l}}}}\text{+}{{{\rarrow{v}}}_{{{\phi }_{n,l}}}}\hat{\lambda }y_{n,l}}{1\text{+}{{{\rarrow{v}}}_{{{\phi }_{n,l}}}}\hat{\lambda }},\frac{{{{\rarrow{v}}}_{{{\phi }_{n,l}}}}}{1\text{+}{{{\rarrow{v}}}_{{{\phi }_{n,l}}}}\hat{\lambda }} \right) \\
& \triangleq \mathcal{CN}\left( \phi _{n,l};\hat{\phi }_{n,l},{{v}_{{{\phi }_{n,l}}}} \right)
\end{salign}
Then, the forward message $ {{I}_{f_{\tilde{y} _{\scriptscriptstyle n,l}}\to {\lambda }}}( {\lambda } ) $ can be calculated by
\begin{salign}
{{I}_{f_{\tilde{y} _{\scriptscriptstyle n,l}}\to {\lambda }}}\left( {\lambda } \right)
\notag &=\exp\!\! \left\{ \!\int \!{\ln \left[ f_{\tilde{y} _{\scriptscriptstyle n,l}}\left( \phi _{n,l},{\lambda }\right)\right]b\big( \phi _{n,l}\big)d\phi _{n,l}} \right\} \\
& \varpropto {\lambda }\exp \left\{ -{\lambda }{{\left| y_{n,l}- \hat{\phi} _{n,l} \right|}^{2}} \right\}
\end{salign}
Thus, with the prior of noise precision $ {{f}_{\lambda }} \big( \lambda \big)$, the belief of $ \lambda $ can be updated by
\begin{salign}\label{eq:blfLmd}
b\left( \lambda  \right) & \varpropto \prod\limits_{l=1}^{L}{\prod\limits_{n=1}^{N}{{{I}_{f_{y_{n,l}}\to \lambda }}\left( \lambda  \right)}}{{f}_{\lambda }}\left( \lambda  \right) \\
\notag & ={\lambda^{NL-1}}\exp \left\{ -\lambda \sum\limits_{l=1}^{L}{\sum\limits_{n=1}^{N} {{\left| y_{n,l}-\hat{\phi} _{n,l} \right|}^{2}} } \right\},
\end{salign}
and the noise  precision is given by
\begin{equation} \label{eq:lambda}
\hat{\lambda} ={{\big\langle \lambda  \big\rangle }_{b\left( \lambda  \right)}} \\
= \frac{NL}{\sum\limits_{L=1}^{L}{\sum\limits_{n=1}^{N} {{\left| y_{n,l}-\hat{\phi} _{n,l} \right|}^{2}} }}.
\end{equation}
The backward message $ {{I}_{f_{\tilde{y} _{\scriptscriptstyle n,l}}\to \phi _{n,l}}}\big( \phi _{n,l} \big) $ can be expressed as
\begin{salign}\label{eq:fy2phi}
{{I}_{f_{\tilde{y} _{\scriptscriptstyle n,l}}\to \phi _{n,l}}}\left( \phi _{n,l} \right)
\notag & \!=\!\exp \!\left\{ \! \int \! {\ln \left[ f_{\tilde{y} _{\scriptscriptstyle n,l}}\left( \phi _{n,l},{\lambda } \right) \right] b( {\lambda } )d{\lambda }} \right\} \\
& \varpropto \mathcal{CN}\left( \phi _{n,l};y_{n,l},{1}/{\hat{\lambda }}\; \right).
\end{salign}

With  {\small $\big\lbrace  I_{f_{\psi _{\scriptscriptstyle n,k',l}}\to \psi _{n,k',l}} \left(\psi _{n,k',l} \right)  ,\forall k' \neq k \big\rbrace $}  in (\ref{eq:fpsi2psi}), the backward message {\small ${{I}_{\scriptscriptstyle \psi _{n,k,l}\to f_{\psi _{\scriptscriptstyle n,k,l}}}} \left(\psi _{n,k,l}\right)$} is calculated as
\begin{salign}\label{eq:psi2fpsi}
\notag &{{I}_{\psi _{n,k,l}\to f_{\psi _{\scriptscriptstyle n,k,l}}}}\left( \psi _{n,k,l} \right) \\
\notag & = {\Big\langle  {f_{\phi _{\scriptscriptstyle n,l}}\big( \phi _{n,l},\psi _{n,k,l},\forall k \big)} \Big\rangle}_{{{I}_{f_{\tilde{y} _{\scriptscriptstyle n,l}}\to \phi _{n,l}}}\left( \phi _{n,l} \right)  \prod\limits_{{k}'\ne k}  I_{f_{\psi _{\scriptscriptstyle n,k',l}}\to \psi _{n,k',l}} \left(\psi _{n,k',l} \right)} \\
\notag & =\mathcal{CN} \Big( \phi _{n,l};y_{n,l} - \sum\limits_{{k}'\ne k} \!\! {{{{\rarrow{m}}}_{{{\psi }_{n,k',l}}}}},{1}/{\hat{\lambda }} + \sum\limits_{{k}'\ne k} \!{{{{\rarrow{v}}}_{{{\psi }_{n,k',l}}}}} \Big) \\
& \triangleq \mathcal{CN}\left( \psi _{n,k,l};{{{\larrow{m}}}_{{{\psi }_{n,k,l}}}},{{{\larrow{v}}}_{{{\psi }_{n,k,l}}}} \right) .
\end{salign}
Then, with the message {\small $ {{I}_{x_{\scriptscriptstyle k,l}\to f_{\psi _{\scriptscriptstyle n,k,l}}}} \big( x_{\scriptscriptstyle k,l} \big) $} in (\ref{eq:x2fpsi}), the forward message {\small $ {{I}_{f_{\psi _{\scriptscriptstyle n,k,l}}\to {h_{n,k}}}}\big( {h_{n,k}} \big) $} is given by
\begin{salign}
{{I}_{f_{\psi _{\scriptscriptstyle n,k,l}}\to {h_{n,k}}}} \!\! \left( {h_{n,k}} \right) \!\!
\notag &  = {\Big\langle \! f_{\psi _{\scriptscriptstyle n,k,l}} \big(\psi _{n,k,l},{h_{n,k}},x_{\scriptscriptstyle k,l}\big) \! \Big\rangle}_{  {{I}_{x_{\scriptscriptstyle k,l}\to f_{\psi _{\scriptscriptstyle n,k,l}}}} \left( x_{\scriptscriptstyle k,l}\right) {{I}_{\psi _{n,k,l}\to f_{\psi _{\scriptscriptstyle n,k,l}}}}\left( \psi _{n,k,l} \right) } \\
& =\sum\limits_{q\in \mathcal{X}}{\gamma^{q}_{\scriptscriptstyle k,l}\mathcal{CN}\left( {h_{n,k}}q;{{{\larrow{m}}}_{{{\psi }_{n,k,l}}}},{{{\larrow{v}}}_{{{\psi }_{n,k,l}}}} \right)} .
\end{salign}
which can be approximated to be Gaussian by
\begin{salign}\label{eq:fpsi2h}
{{I}_{f_{\psi _{\scriptscriptstyle n,k,l}}\to {h_{n,k}}}}\left( {h_{n,k}} \right)
\notag &  \approx \text{Proj}_{G}\left\{ {{I}_{f_{\psi _{\scriptscriptstyle n,k,l}}\to {h_{n,k}}}}\left( {h_{n,k}} \right) \right\} \\
& \triangleq \mathcal{CN}\left( {h_{n,k}};{{{\overset{\scriptscriptstyle\twoheadrightarrow}{m}}}_{{h_{n,k}}}},{{{\overset{\scriptscriptstyle\twoheadrightarrow}{v}}}_{{h_{n,k}}}} \right),
\end{salign}
where
\begin{salign}\label{eq:Mfpsi2h}
{{{\overset{\scriptscriptstyle\twoheadrightarrow}{m}}}_{{h_{n,k}}}}
 = {\big\langle  {h_{n,k}} \big\rangle }_{{{I}_{f_{\psi _{\scriptscriptstyle n,k,l}}\to {h_{n,k}}}}\left( {h_{n,k}} \right)}
 =\frac{\sum\limits_{q\in \mathsf{\mathcal{X}}}{{\gamma^{q}_{\scriptscriptstyle k,l}}/{{{\left| q \right|}^{2}}\cdot }\;{{{{\larrow{m}}}_{{{\psi }_{n,k,l}}}}}/{q} }}{\sum\limits_{q\in \mathsf{\mathcal{X}}}{{\gamma^{q}_{\scriptscriptstyle k,l}}/{{{\left| q \right|}^{2}}}}}  ,
\end{salign}

\begin{algorithm} [t]
	\caption{ BP-MF based MUD joint with UAD and CE}
	\label{alg:H-MP}
	{\bf Input:}
	{\small $ \textbf{Y},  p\left( \lambda  \right),   \left\lbrace p\left( {z}_{\scriptscriptstyle k}\right) \right\rbrace,   \left\lbrace p\left( {g_{n,k}}\right) \right\rbrace,   \left\lbrace p\left(  x_{\scriptscriptstyle k,l} \right) \right\rbrace$}. The initial values {\small $ \hat{\lambda}, \big\lbrace {{\rarrow{m}}}_{{h_{n,k}}}, {{\rarrow{v}}}_{{h_{n,k}}} \big\rbrace $ and $ \big\lbrace\hat x_{\scriptscriptstyle k,l} \big\rbrace$}, which are  provided by   {Algorithm} \ref{preA2}.\vspace{1.2mm}\\
	1: \textbf{for} {\small $ i=1:N_{O_{itr}} $} (Outer iteration)\\
	2: \hspace*{0.15 in} $ \forall k,u $:  update {\small $ \beta _{k}^{u} $} by (\ref{eq:BetaKU}).\\
	3: \hspace*{0.15 in} $ \forall n,k $:  update {\small $ {{{\hat{h}}}_{n,k}} $} and {\small $ v_{{h_{n,k}}} $} by (\ref{eq:epMh}) and (\ref{eq:epVh}).\\
	4: \hspace*{0.15 in} $ \forall n,k $:  update {\small $ {{{\larrow{m}}}_{{h_{n,k}}}} $} and {\small $ {{{\larrow{v}}}_{{h_{n,k}}}} $} by (\ref{eq:MVfh2h}).\\
	5: \hspace*{0.15 in} \textbf{for} {\small $ j=1:N_{I_{itr}} $} (Inner iteration)\\	
	6: \hspace*{0.3 in} $ \forall n,k $:  update {\small $ {{\overset{\scriptscriptstyle\twoheadleftarrow}{m}}_{{h_{n,k}}}} $} and {\small $ {{\overset{\scriptscriptstyle\twoheadleftarrow}{v}}_{{h_{n,k}}}} $} by (\ref{eq:MVh2fpsi}).\\	
	7: \hspace*{0.3 in} $ \forall n,k,l $:  update {\small $ {{{\rarrow{m}}}_{{{\psi }_{n,k,l}}}} $} and {\small $ {{{\rarrow{v}}}_{{{\psi }_{n,k,l}}}} $} by (\ref{eq:Mfpsi2psiA2}) and (\ref{eq:Vfpsi2psiA2}).\\	
	8: \hspace*{0.3 in} $ \forall n,l $:  update {\small $ {{{\rarrow{m}}}_{{{\phi }_{n,l}}}} $} and {\small $ {{{\rarrow{v}}}_{{{\phi }_{n,l}}}} $} in (\ref{eq:fphi2phi}).\\	
	9: \hspace*{0.3 in} $ \forall n,l $:  update {\small $\hat{\phi }_{n,l} $} in (\ref{blfphi}).\\
	10: \hspace*{0.25 in}   update  {\small $ \hat{\lambda} $} by (\ref{eq:lambda}).\\		
	11: \hspace*{0.25 in} $ \forall n,k,l $:  update {\small $ {{{\larrow{m}}}_{{{\psi }_{n,k,l}}}} $} and {\small $ {{{\larrow{v}}}_{{{\psi }_{n,k,l}}}} $} in (\ref{eq:psi2fpsi}).\\
	12: \hspace*{0.25 in} $ \forall k,l,q $: update {\small $ \gamma^{q}_{\scriptscriptstyle k,l} $}  by (\ref{eq:Gammaqk,l}).\\		
	13: \hspace*{0.25 in} $ \forall n,k $:  update {\small $ {{\overset{\scriptscriptstyle\twoheadrightarrow}{m}}_{{h_{n,k}}}} $} and {\small $ {{\overset{\scriptscriptstyle\twoheadrightarrow}{v}}_{{h_{n,k}}}} $} by (\ref{eq:Mfpsi2h}) and (\ref{eq:Vfpsi2h}).\\
	14: \hspace*{0.25 in} $ \forall n,k $:  update {\small $ {{{\rarrow{m}}}_{{h_{n,k}}}} $} and {\small $  {{{\rarrow{v}}}_{{h_{n,k}}}}$} by (\ref{eq:MVh2fh2}).\\	
	15: \hspace*{0.25 in} $ \forall n,k $:  update {\small $ \hat h_{n,k}$} and {\small $ v_{{h_{n,k}}} $} by (\ref{eq:MVh}).\\		
	16: \hspace*{0.1 in} \textbf{end}\\		
	17: \textbf{end}\\
	{\bf Output:} Posterior distributions:
	{\small $  b\left(z_{k}\right)= \sum\limits_{u\in \mathsf{\mathcal{U}}}{\beta _{k}^{u}\delta \left( {{z}_{\scriptscriptstyle k}}-u \right)}, \forall k;$}\\
	\hspace*{0.55 in}{\small $  \hat h_{n,k},\forall n,k; \qquad  {{b}}\left( x_{\scriptscriptstyle k,l} \right) = \sum\limits_{q\in \mathsf{\mathcal{X}}}{\beta _{\scriptscriptstyle k,l}^{q}\delta \left( x_{\scriptscriptstyle k,l}-q \right)}, \forall k,l.$}
\end{algorithm}

\begin{salign}\label{eq:Vfpsi2h}
{{{\overset{\scriptscriptstyle\twoheadrightarrow}{v}}}_{{h_{n,k}}}}
\notag &= {{\text{Var}\big[ h_{n,k}  \big] }_{{{I}_{f_{\psi _{\scriptscriptstyle n,k,l}}\to {h_{n,k}}}}\left( {h_{n,k}} \right)}} \\
& =\frac{\sum\limits_{q\in \mathsf{\mathcal{X}}}{{\gamma^{q}_{\scriptscriptstyle k,l}}/{{{\left| q \right|}^{4}}\cdot }\;\Big( {{\big| {{{\larrow{m}}}_{{{\psi }_{n,k,l}}}} \big|}^{2}}+{{{\larrow{v}}}_{{{\psi }_{n,k,l}}}} \Big)}}{\sum\limits_{q\in \mathsf{\mathcal{X}}}{{\gamma^{q}_{\scriptscriptstyle k,l}}/{{{\left| q \right|}^{2}}}\;}}-{{\left| {{{\overset{\scriptscriptstyle\twoheadrightarrow}{m}}}_{{h_{n,k}}}} \right|}^{2}}  .
\end{salign}
Thus, the forward message $ {{I}_{{h_{n,k}}\to {{f}_{{h_{n,k}}}}}}\left( {h_{n,k}} \right) $ , which is used in Part (i), becomes
\begin{salign}\label{eq:htofh2}
\notag {{I}_{{h_{n,k}}\to {{f}_{{h_{n,k}}}}}}\left( {h_{n,k}} \right) & =\prod\limits_{n=1}^{N}{ {{I}_{f_{\psi _{\scriptscriptstyle n,k,l}}\to {h_{n,k}}}}\left( {h_{n,k}} \right) } \\
& \varpropto \mathcal{CN}\left( {h_{n,k}};{{{\rarrow{m}}}_{{h_{n,k}}}},{{{\rarrow{v}}}_{{h_{n,k}}}} \right),
\end{salign}
where
\begin{salign}\label{eq:MVh2fh2}
{{{\rarrow{v}}}_{{h_{n,k}}}} ={{\left( \sum\limits_{n=1}^{N}{{{{\overset{\scriptscriptstyle\twoheadrightarrow}{v}}_{{h_{n,k}}}}}^{-1}} \right)}^{-1}}, \quad
{{{\rarrow{m}}}_{{h_{n,k}}}}  = \Bigg( \sum\limits_{n=1}^{N}{\frac{{{\overset{\scriptscriptstyle\twoheadrightarrow}{m}}_{{h_{n,k}}}}}{{{\overset{\scriptscriptstyle\twoheadrightarrow}{v}}_{{h_{n,k}}}}}} \Bigg){{{\rarrow{v}}}_{{h_{n,k}}}}.
\end{salign}
Then, with {\small $ I^{\scriptscriptstyle EP}_{{{f}_{{h_{n,k}}}}\to {h_{n,k}}} \big( {h_{n,k}} \big) \varpropto \mathcal{CN}\big( {h_{n,k}};{{{\larrow{m}}}_{{h_{n,k}}}},{{{\larrow{v}}}_{{h_{n,k}}}} \big) $}  from Part (i) in (\ref{eq:EPfh2h}), the belief {\small $ b\left({h_{n,k}}\right) $} can be updated in the same way as (\ref{eq:blfHa1})-(\ref{eq:MVh}).


\begin{algorithm} [t]
	\caption{ Pre-processor used in Algorithm 2}
	\label{preA2}
	{\bf Input:}
	{\small $ \textbf{Y},  p\left( \lambda  \right),   \left\lbrace p\left( {z}_{\scriptscriptstyle k}\right) \right\rbrace,   \left\lbrace p\left( {g_{n,k}}\right) \right\rbrace,   \left\lbrace p\left(  x_{\scriptscriptstyle k,l} \right) \right\rbrace.$ } \\
	\textbf{Initialize:} {\small $ \hat{\lambda}; \big\lbrace {{\rarrow{m}}}_{{h_{n,k}}}, {{\rarrow{v}}}_{{h_{n,k}}} \big\rbrace  ; \  \forall k,l, \hat x_{\scriptscriptstyle k,l}=q, q \in \mathcal{X}$}\\  	
	1: \textbf{for} {\small $ i=1:N_{O_{itr}} $} (Outer iteration) \\	
	2: \hspace*{0.15 in} $ \forall k,u $:  update {\small $ \beta _{k}^{u} $} by (\ref{eq:BetaKU}). \\	
	3: \hspace*{0.15 in} $ \forall n,k $:  update {\small $ {{{\hat{h}}}_{n,k}} $} and {\small $ v_{{h_{n,k}}} $} by (\ref{eq:epMh}) and (\ref{eq:epVh}). \\	
	4: \hspace*{0.15 in} $ \forall n,k $:  update {\small $ {{{\larrow{m}}}_{{h_{n,k}}}} $} and {\small $ {{{\larrow{v}}}_{{h_{n,k}}}} $} by (\ref{eq:MVfh2h}). \\
	5: \hspace*{0.15 in} \textbf{for} {\small $ j=1:N_{I_{itr}} $} (Inner iteration)\\
	6: \hspace*{0.3 in} $ \forall k,l $:  update {\small $ {{\overset{\scriptscriptstyle\twoheadrightarrow}{m}}_{{{x}_{\scriptscriptstyle k,l}}}} $} and {\small $ {{\overset{\scriptscriptstyle\twoheadrightarrow}{v}}_{{{x}_{\scriptscriptstyle k,l}}}} $} by (\ref{eq:MVfpsi2xinit}). \\
	7: \hspace*{0.3 in}  $ \forall n,k $:  update {\small $ {{\overset{\scriptscriptstyle\twoheadleftarrow}{m}}_{{h_{n,k}}}} $} and {\small $ {{\overset{\scriptscriptstyle\twoheadleftarrow}{v}}_{{h_{n,k}}}} $} by (\ref{eq:MVh2fpsi}).\\
	8: \hspace*{0.3 in} $ \forall n,k,l $:  update {\small $ {{{\rarrow{m}}}_{{{\psi }_{n,k,l}}}} $} and {\small $ {{{\rarrow{v}}}_{{{\psi }_{n,k,l}}}} $} by (\ref{eq:Mfpsi2psi}) and (\ref{eq:Vfpsi2psi}).\\	
	9: \hspace*{0.3  in} $ \forall n,l $:  update {\small $ {{{\rarrow{m}}}_{{{\phi }_{n,l}}}} $} and {\small $ {{{\rarrow{v}}}_{{{\phi }_{n,l}}}} $} in (\ref{eq:fphi2phi}).\\
	10: \hspace*{0.25  in} $ \forall n,l $:  update {\small $\hat{\phi }_{n,l} $} in (\ref{blfphi}).\\
	11: \hspace*{0.25 in}  update  {\small $ \hat{\lambda} $} by (\ref{eq:lambda}).\\	
	12: \hspace*{0.25 in} $ \forall n,k,l $:  update {\small $ {{{\larrow{m}}}_{{{\psi }_{n,k,l}}}} $} and {\small $ {{{\larrow{v}}}_{{{\psi }_{n,k,l}}}} $} in (\ref{eq:psi2fpsi}).	\\
	13: \hspace*{0.25 in} $ \forall k,l $:  update {\small $ {{\overset{\scriptscriptstyle\twoheadleftarrow}{m}}_{{{x}_{\scriptscriptstyle k,l}}}} $} and {\small $ {{\overset{\scriptscriptstyle\twoheadleftarrow}{v}}_{{{x}_{\scriptscriptstyle k,l}}}} $} by (\ref{eq:MVfy2xinit}).	\\
	14: \hspace*{0.25 in}  $ \forall k,l,q $:  update {\small $ \beta _{\scriptscriptstyle k,l}^{q} $} by (\ref{eq:Beta_klq}).	\\
	15: \hspace*{0.25 in} $ \forall k,l $:  update {\small $ \hat x_{\scriptscriptstyle k,l} $} and {\small $ {{v}_{{{x}_{\scriptscriptstyle k,l}}}} $} by (\ref{eq:Mx}) and (\ref{eq:Vx}).\\	
	16: \hspace*{0.25 in} $ \forall n,k $:  update {\small $ {{\overset{\scriptscriptstyle\twoheadrightarrow}{m}}_{{h_{n,k}}}} $} and {\small $ {{\overset{\scriptscriptstyle\twoheadrightarrow}{v}}_{{h_{n,k}}}} $} by (\ref{eq:MVfy2hinit}).	\\
	17: \hspace*{0.25 in} $ \forall n,k $:  update {\small $ {{{\rarrow{m}}}_{{h_{n,k}}}} $} and {\small $  {{{\rarrow{v}}}_{{h_{n,k}}}}$} by (\ref{eq:MVh2fh2}).\\
	18: \hspace*{0.25 in} $ \forall n,k $:  update {\small $ {{{\hat{h}}}_{n,k}} $} and {\small $ v_{{h_{n,k}}} $} by (\ref{eq:MVh}).\\	
	19: \hspace*{0.1 in} \textbf{end} \\		
	20: \textbf{end}\\ 	
\end{algorithm}

It is worth mentioning that, compared to the pure MF based method in Section III.C.(1), the BP-MF based method improve the system performance  at the cost of increased computational complexity.
As the message computations need to be executed for a certain number of  iterations, it is desirable to use the least number of iterations in the BP-MF method. Our strategy is to design a low-complexity pre-processor for providing initial messages for the BP-MF  method, so that the BP-MF method can converge rapidly while achieving good performance. The low-complexity pre-processor is elaborated in the following.


Inspired by the pure MF based method,  the belief of $ x_{\scriptscriptstyle k,l} $ can be  approximated to be Gaussian, given in (\ref{eq:blfXInit}). With {\small $ {{I}_{f_{\psi _{\scriptscriptstyle n,k,l}}\to  x_{\scriptscriptstyle k,l} }}\left( x_{\scriptscriptstyle k,l} \right) $}  given in (\ref{eq:fpsi2xInit}), the forward message {\small $ {{I}_{x_{\scriptscriptstyle k,l}\to f_{\psi _{\scriptscriptstyle n,k,l}}}} \left( x_{\scriptscriptstyle k,l} \right) $} is calculated by
\begin{salign}
	{{I}_{ x_{\scriptscriptstyle k,l} \to f_{\psi _{\scriptscriptstyle n,k,l}}}}\left( x_{\scriptscriptstyle k,l} \right)   =\frac{b\left( x_{\scriptscriptstyle k,l} \right)}{{{I}_{f_{\psi _{\scriptscriptstyle n,k,l}}\to  x_{\scriptscriptstyle k,l} }}\left( x_{\scriptscriptstyle k,l} \right)}
	\varpropto \mathcal{CN}\left( x_{\scriptscriptstyle k,l} ; {{{\overset{\scriptscriptstyle\twoheadrightarrow}{m}}}_{{{x}_{\scriptscriptstyle k,l}}}},{{{\overset{\scriptscriptstyle\twoheadrightarrow}{v}}}_{{{x}_{\scriptscriptstyle k,l}}}} \right),
\end{salign}
where
\begin{salign} \label{eq:MVfpsi2xinit}
	{{\overset{\scriptscriptstyle\twoheadrightarrow}{v}}_{x_{\scriptscriptstyle k,l} }}={\Bigg( \frac{1}{{{v}_{x_{\scriptscriptstyle k,l} }}} - \frac{1}{{{\overset{\scriptscriptstyle\twoheadleftarrow}{v}}_{x_{\scriptscriptstyle k,l} }}}  \Bigg)^{-1}}, \quad
	{{\overset{\scriptscriptstyle\twoheadrightarrow}{m}}_{ x_{\scriptscriptstyle k,l}}}=\Bigg( \frac{{{{\hat{x}}}_{ \scriptscriptstyle kl }}}{{{v}_{x_{\scriptscriptstyle k,l} }}} - \frac{{{\overset{\scriptscriptstyle\twoheadleftarrow}{m}}_{{h_{n,k}}}}}{{{\overset{\scriptscriptstyle\twoheadleftarrow}{v}}_{x_{\scriptscriptstyle k,l} }}}  \Bigg) {{\overset{\scriptscriptstyle\twoheadrightarrow}{v}}_{ x_{\scriptscriptstyle k,l}}}.
\end{salign}

With {\small $ {{I}_{f_{\psi _{\scriptscriptstyle n,k,l}}\to {h_{n,k}}}} \big( {h_{n,k}} \big)$} given in (\ref{eq:fpsi2hInit}),   {\small $ b\left( h_{n,k}\right)  $} can be calculated in the same way to (\ref{eq:blfHa1})-(\ref{eq:MVh}) and (\ref{eq:htofh2})-(\ref{eq:MVh2fh2}). Then, {\small $ {{I}_{{h_{n,k}} \to f_{\psi _{\scriptscriptstyle n,k,l}}}} \big( {h_{n,k}} \big)  \propto \mathcal{CN}\big( {h_{n,k}};{{\overset{\scriptscriptstyle\twoheadleftarrow}{m}}_{{h_{n,k}}}},{{\overset{\scriptscriptstyle\twoheadleftarrow}{v}}_{{h_{n,k}}}} \big)$} is calculated in the same way to (\ref{eq:h2fpsi}).
Thus, the forward message {\small $ {{I}_{f_{\psi _{\scriptscriptstyle n,k,l}}\to \psi _{n,k,l}}} \big(\psi _{n,k,l}\big) $}  can be approximated to Gaussian be by
\begin{salign}\label{eq:fpsi2psiInit}
\notag & I_{f_{\psi _{\scriptscriptstyle n,k,l}}\to \psi _{n,k,l}}\left( \psi _{n,k,l} \right)\\
\notag &\approx\text{Proj}_{G}\left\{ {\Big\langle  f_{\psi _{\scriptscriptstyle n,k,l}} \big(\psi _{n,k,l},{h_{n,k}},x_{\scriptscriptstyle k,l}\big) \Big\rangle}_{ {{I}_{ x_{\scriptscriptstyle k,l} \to f_{\psi _{\scriptscriptstyle n,k,l}}  }} ( x_{\scriptscriptstyle k,l})  {{I}_{{h_{n,k}} \to f_{\psi _{\scriptscriptstyle n,k,l}}}} ({h_{n,k}})} \right\} \\
& \triangleq \mathcal{CN}\left( \psi _{n,k,l};{{{\rarrow{m}}}_{{{\psi }_{n,k,l}}}},{{{\rarrow{v}}}_{{{\psi }_{n,k,l}}}} \right),
\end{salign}
where the mean and variance of $  {\psi_{n,k,l}} $ are given by
\begin{salign}
\label{eq:Mfpsi2psi} &{{{\rarrow{m}}}_{{{\psi }_{n,k,l}}}} = {{\overset{\scriptscriptstyle\twoheadleftarrow}{m}}_{{h_{n,k}}}}  {{{\overset{\scriptscriptstyle\twoheadrightarrow}{m}}}_{{{x}_{\scriptscriptstyle k,l}}}}  ,  \\
\label{eq:Vfpsi2psi}&{{{\rarrow{v}}}_{{{\psi }_{n,k,l}}}}= {{\overset{\scriptscriptstyle\twoheadleftarrow}{v}}_{{h_{n,k}}}}  \big| {{{\overset{\scriptscriptstyle\twoheadrightarrow}{m}}}_{{{x}_{\scriptscriptstyle k,l}}}} \big|^2 + {{{\overset{\scriptscriptstyle\twoheadrightarrow}{v}}}_{{{x}_{\scriptscriptstyle k,l}}}} \big|  {{\overset{\scriptscriptstyle\twoheadleftarrow}{m}}_{{h_{n,k}}}} \big|^2+ {{\overset{\scriptscriptstyle\twoheadleftarrow}{v}}_{{h_{n,k}}}} {{{\overset{\scriptscriptstyle\twoheadrightarrow}{v}}}_{{{x}_{\scriptscriptstyle k,l}}}}   .
\end{salign}

With {\small $ {{I}_{\psi _{n,k,l}\to f_{\psi _{\scriptscriptstyle n,k,l}}}} ( \psi _{n,k,l}  )\propto $}  {\small  $ \mathcal{CN} ( \psi _{n,k,l};{{{\larrow{m}}}_{{{\psi }_{n,k,l}}}},{{{\larrow{v}}}_{{{\psi }_{n,k,l}}}}  ) $} in (\ref{eq:psi2fpsi}), the function node {\small $ f_{\psi _{\scriptscriptstyle n,k,l}}  (\psi _{n,k,l},{h_{n,k}},x_{\scriptscriptstyle k,l} ) $} can be integrated with respect to  {\small $ \psi _{n,k,l} $}, leading to  a new function
 {\small $ \hat{f}_{\psi _{\scriptscriptstyle n,k,l}}  ({h_{n,k}},x_{\scriptscriptstyle k,l} ) $},
\begin{salign}
\hat{f}_{\psi _{\scriptscriptstyle n,k,l}} \big({h_{n,k}},x_{\scriptscriptstyle k,l}\big)
\notag & = {\Big\langle  f_{\psi _{\scriptscriptstyle n,k,l}} \big(\psi _{n,k,l},{h_{n,k}},x_{\scriptscriptstyle k,l}\big) \Big\rangle}_{{{I}_{\psi _{n,k,l}\to f_{\psi _{\scriptscriptstyle n,k,l}}}}\left( \psi _{n,k,l} \right)}\\
& = \mathcal{CN}\left({h_{n,k}} x_{\scriptscriptstyle k,l} ;{{{\larrow{m}}}_{{{\psi }_{n,k,l}}}},{{{\larrow{v}}}_{{{\psi }_{n,k,l}}}} \right),
\end{salign}
Then, MF is performed to calculate the forward message {\small $ {{I}_{f_{\psi _{\scriptscriptstyle n,k,l}}\to {h_{n,k}}}} \big( {h_{n,k}} \big) $} and   the backward message {\small $ {{I}_{f_{\psi _{\scriptscriptstyle n,k,l}}\to x_{\scriptscriptstyle k,l}}}( x_{\scriptscriptstyle k,l}) $ } as
\begin{salign}\label{eq:fpsi2hInit}
	{{I}_{f_{\psi _{\scriptscriptstyle n,k,l}}\to {h_{n,k}}}}\left( {h_{n,k}} \right)
	\notag & = \exp \left\{ \int \ln \left[ \hat{f}_{\psi _{\scriptscriptstyle n,k,l}} \big({h_{n,k}},x_{\scriptscriptstyle k,l}\big) \right]b\left( x_{\scriptscriptstyle k,l} \right) d x_{\scriptscriptstyle k,l} \right\}\\
	& \varpropto \mathcal{CN}\left( {h_{n,k}};{{{\overset{\scriptscriptstyle\twoheadrightarrow}{m}}}_{{h_{n,k}}}},{{{\overset{\scriptscriptstyle\twoheadrightarrow}{v}}}_{{h_{n,k}}}} \right)
\end{salign}
with
\begin{salign} \label{eq:MVfy2hinit}
	{{{\overset{\scriptscriptstyle\twoheadrightarrow}{m}}}_{{h_{n,k}}}}
	= \frac{ {\hat x_{\scriptscriptstyle k,l}}^{H} {{{\larrow{m}}}_{{{\psi }_{n,k,l}}}}}{{{\left| \hat x_{\scriptscriptstyle k,l} \right|}^{2}} + {{v}_{{{x}_{\scriptscriptstyle k,l}}}}}, \quad
	{{{\overset{\scriptscriptstyle\twoheadrightarrow}{v}}}_{{h_{n,k}}}}
	= \frac{{{{\larrow{v}}}_{{{\psi }_{n,k,l}}}} }{ {{\big| \hat x_{\scriptscriptstyle k,l} \big|}^{2}}+{{v}_{{{x}_{\scriptscriptstyle k,l}}}} },
\end{salign}
and
\begin{salign}\label{eq:fpsi2xInit}
	{{I}_{f_{\psi _{\scriptscriptstyle n,k,l}}\to x_{\scriptscriptstyle k,l}}}( x_{\scriptscriptstyle k,l})
	\notag &=\exp \left\{ \int \ln \left[ \hat{f}_{\psi _{\scriptscriptstyle n,k,l}} \big({h_{n,k}},x_{\scriptscriptstyle k,l}\big) \right]b\left( {h_{n,k}} \right) d{h_{n,k}} \right\}\\
	& \varpropto \mathcal{CN}\left( x_{\scriptscriptstyle k,l}; {{{\overset{\scriptscriptstyle\twoheadleftarrow}{m}}}_{{{x}_{\scriptscriptstyle k,l}}}},{{{\overset{\scriptscriptstyle\twoheadleftarrow}{v}}}_{{{x}_{\scriptscriptstyle k,l}}}} \right)
\end{salign}
with
\begin{salign}\label{eq:MVfy2xinit}
	{{\overset{\scriptscriptstyle\twoheadleftarrow}{m}}_{{{x}_{\scriptscriptstyle k,l}}}}
	=\frac{ {{{\hat{h}}}_{n,k}}^{H} {{{\larrow{m}}}_{{{\psi }_{n,k,l}}}}}{{{\left| {{{\hat{h}}}_{n,k}} \right|}^{2}}+{{v}_{{h_{n,k}}}}}, \quad
	{{\overset{\scriptscriptstyle\twoheadleftarrow}{v}}_{{{x}_{\scriptscriptstyle k,l}}}}
	= \frac{{{{\larrow{v}}}_{{{\psi }_{n,k,l}}}} }{ {{\big| {{{\hat{h}}}_{n,k}} \big|}^{2}}+{{v}_{{h_{n,k}}}}}.
\end{salign}
Thus, the belief  of $ x_{\scriptscriptstyle k,l}$ can be updated and approximated to be Gaussian by
\begin{salign}\label{eq:blfXInit}
\notag {{b}}\left( x_{\scriptscriptstyle k,l} \right) & \approx \text{Proj}_{G}\left\{ f_{x_{\scriptscriptstyle k,l}}\left( x_{\scriptscriptstyle k,l} \right)\prod\limits_{n=1}^{N}{ {{I}_{f_{\psi _{\scriptscriptstyle n,k,l}}\to x_{\scriptscriptstyle k,l}}} \left( x_{\scriptscriptstyle k,l} \right)} \right\} \\
& \triangleq \mathcal{CN}\left( x_{\scriptscriptstyle k,l};\hat x_{\scriptscriptstyle k,l},{{v}_{{{x}_{\scriptscriptstyle k,l}}}} \right) ,
\end{salign}
where $ \hat x_{\scriptscriptstyle k,l} $ and $  {{v}_{{{x}_{\scriptscriptstyle k,l}}}} $ can be calculated in a similar way to (\ref{eq:Mx}) and (\ref{eq:Vx}).

The algorithm of message passing based MUD joint with UAD and CE, described in Section III.B and  Section III.C.(2), is summarized in {Algorithm \ref{alg:H-MP}}.

\subsection{Decoding and active user identification }

Both Algorithm \ref{alg:MF-MP} and Algorithm \ref{alg:H-MP}  provide the beliefs of   {\small $\big\lbrace  x_{\scriptscriptstyle k,l}, \forall k,l \big\rbrace $} denoted by {\small $ \big\lbrace {{b}}\left( x_{\scriptscriptstyle k,l} \right), \forall k,l \big\rbrace $}, and the beliefs of {\small $ \big\lbrace z_{k}, \forall k   \big\rbrace $}  denoted by {\small $ \big\lbrace  b\left(z_{k}\right), \forall k \big\rbrace  $}, and {\small $ \big\lbrace \hat h_{n,k},\forall n,k  \big\rbrace $}.
{\small $ \big\lbrace {{b}}\left( x_{\scriptscriptstyle k,l} \right), \forall k,l \big\rbrace $} are used for soft demodulation and decoding.  {\small $ \big\lbrace  b\left(z_{k}\right), \forall k \big\rbrace  $} are used for active user identification, i.e., {\small $ \hat z_{k} = \mathop{\arg\max}_{z_{k}} b\left(z_{k}\right)$} indicates that active user $ k $ employs the LDS sequence $ \textbf{s}_{\hat z_{k}} $.  As each user is allocated a unique LDS sequence, so active user $ k $ is identified.


\begin{table}
	\centering
	\caption{Simulation Parameters}\label{tab2}
	\begin{tabular}{m{4.5cm} c m{2cm}}
		\toprule[1pt]
		\textbf{Parameter} & \textbf{Symbol} & \textbf{Value} \\
		\midrule[0.5pt]
		Number of users & $ U $ & 256 \\
		Number of subcarriers & $ N $ & 128 \\
		Number of active users & $ K $ &  $ 25 $ \\
		Length of symbol sequences & $ L $ & 40 \\
		Number of subcarriers occupied by each user & $ d_c $ & 16 or 32 \\
		\bottomrule[1pt]
	\end{tabular}
\end{table}

\begin{figure}[t]
	\centering
	\includegraphics[width=3.5in]{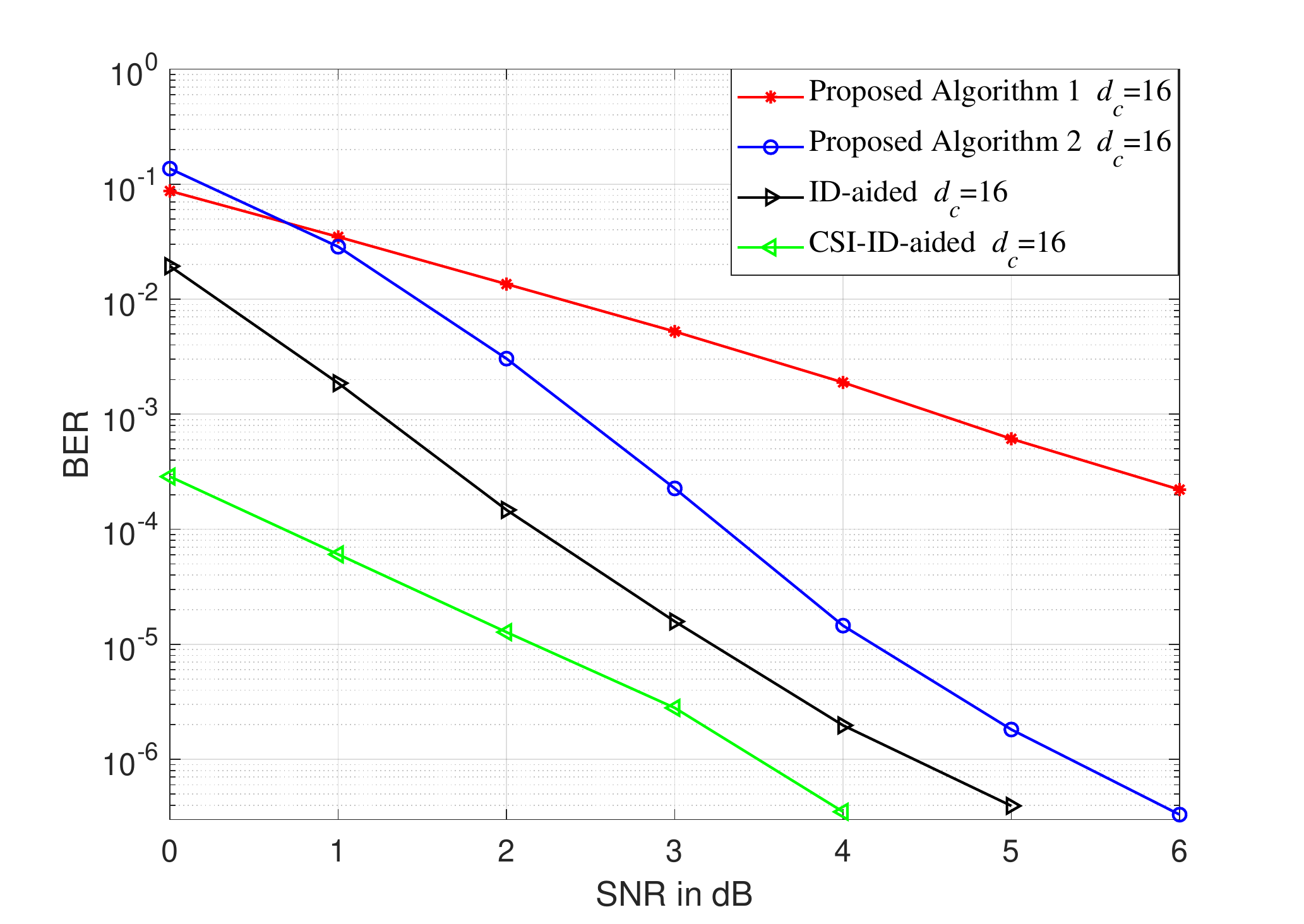}
	\caption{ BER performance comparison. }\label{fig:BER}
\end{figure}

\begin{figure}[t]
	\centering
	\includegraphics[width=3.5in]{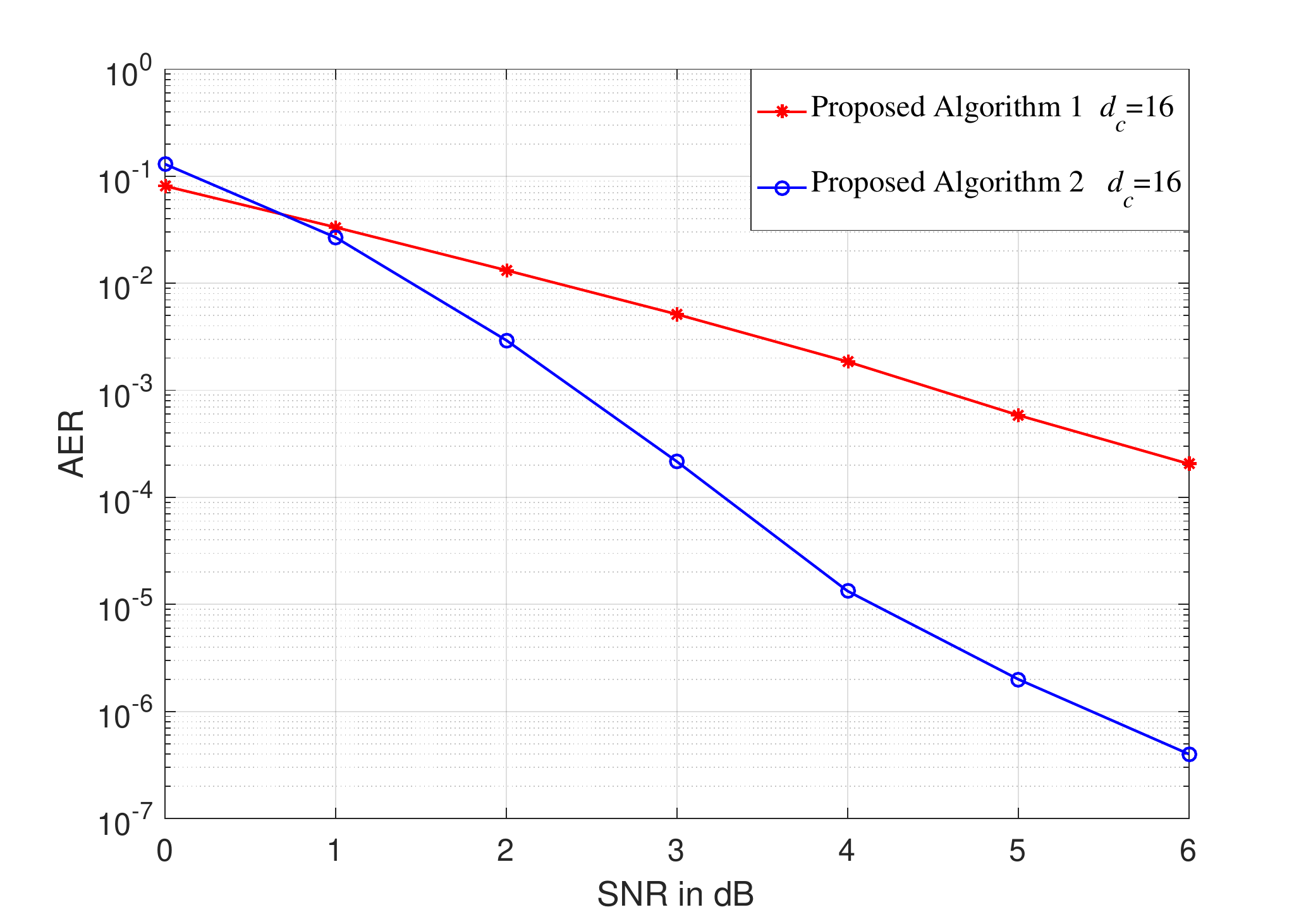}
	\caption{ AER performance comparison. }\label{fig:AER}
\end{figure}

\subsection{Complexity Analysis }
Both {Algorithm \ref{alg:MF-MP}} and  {Algorithm \ref{alg:H-MP}} employ message passing in Part (i) for UAD and CE, and the complexity  is the order of $ O(NKU)+O(NKU)  $ per iteration. The MUD part of  {Algorithm \ref{alg:MF-MP}} has a complexity of $O(NKLQ)+ O(3N^2K) $ per iteration, and that of  {Algorithm \ref{alg:H-MP}} has a complexity of $O(3NKLQ)+ O(2N^2K) $ per iteration. The pre-processor used in  {Algorithm \ref{alg:H-MP}} has a complexity of $ O(NKLQ)+ O(2N^2K) $ per iteration.

\section{Simulation results}
We assume an uplink  LDS-OFDM system  with parameters shown in Table \ref{tab2}. The number of subcarriers $ N=128 $ and the number of users $ U=256 $, i.e., the overloading factor is 2.
The coded modulation scheme RI-TCM in \cite{90RIRCM1999} based on QPSK modulation is employed. We set the number of inner iteration $N_{I_{itr}}=5$ both in  {Algorithm \ref{alg:MF-MP}} and  {Algorithm \ref{alg:H-MP}}. Later, we will show both of the algorithms converge fairly fast, e.g., $N_{O_{itr}}$ is about 10 for  {Algorithm \ref{alg:H-MP}} and 40 for  {Algorithm \ref{alg:MF-MP}}. All the simulation results presented in this section are obtained by averaging over $10^5$ trials.

To the best of our knowledge, the problem of MUD (performed jointly with UAD and CE) of grant-free LDS-OFDM without the use of pilot is investigated in the paper for the first time, which is formulated as a  structured signal estimation problem. Moreover, there are no existing algorithms to solve the formulated problem. So we compare the two proposed algorithms with some corresponding performance bounds. The bit error rate (BER) is used to evaluate the proposed receiver with different algorithms. To examine the performance of user activity detection, we define active user identification error rate (AER) as
\begin{salign}
\notag \text{AER} =\frac{\# \ \text{of active users} - \# \ \text{of active user identified  successfully} }{\# \ \text{of active users}}.
\end{salign}
\normalsize

The BER performance of the proposed schemes is shown in Fig. \ref{fig:BER} for different SNRs, where two additional schemes ID-aided scheme and CSI-ID-aided scheme are also included as benchmarks.
In the ID-aided scheme, we assume that the ID of the active users are perfectly known, but their CSI are not known at the receiver, i.e., the structure of $\textbf{H} $ is known but the values of its non-zero entries are unknown. In the CSI-ID-aided scheme, we assume that both the ID and CSI  of active users are perfectly known, i.e.,  $\textbf{H} $ is  perfectly known, representing the ideal benchmark, and this is served as performance lower bound. It can be seen from the Fig. \ref{fig:BER} that, although having lower complexity,   {Algorithm \ref{alg:MF-MP}}  suffers from significant  performance loss due to the poor efficiency in dealing with observation factors. In contrast, with combined BP and MF to handle the observation factors,  {Algorithm \ref{alg:H-MP}} can improve the performance significantly, with about a 1 dB away from the ID-aided scheme and 2 dB away from the ideal CSI-ID-aided scheme at relatively high SNR range.

\begin{figure}[t]
	\centering
	\includegraphics[width=3.5in]{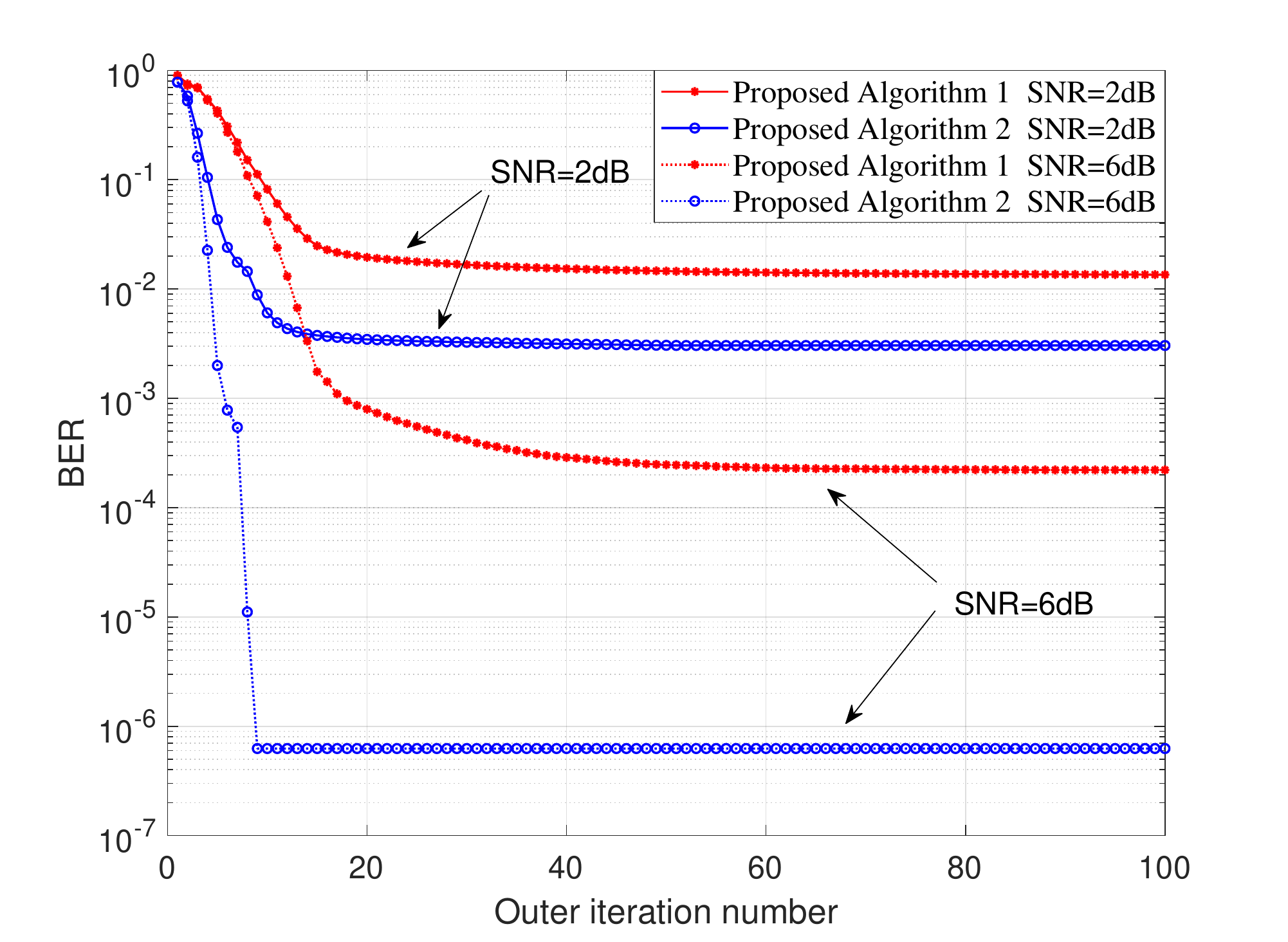}
	\caption{ Convergence of BER. }\label{fig:BERitr}
\end{figure}

\begin{figure}[t]
	\centering
	\includegraphics[width=3.5in]{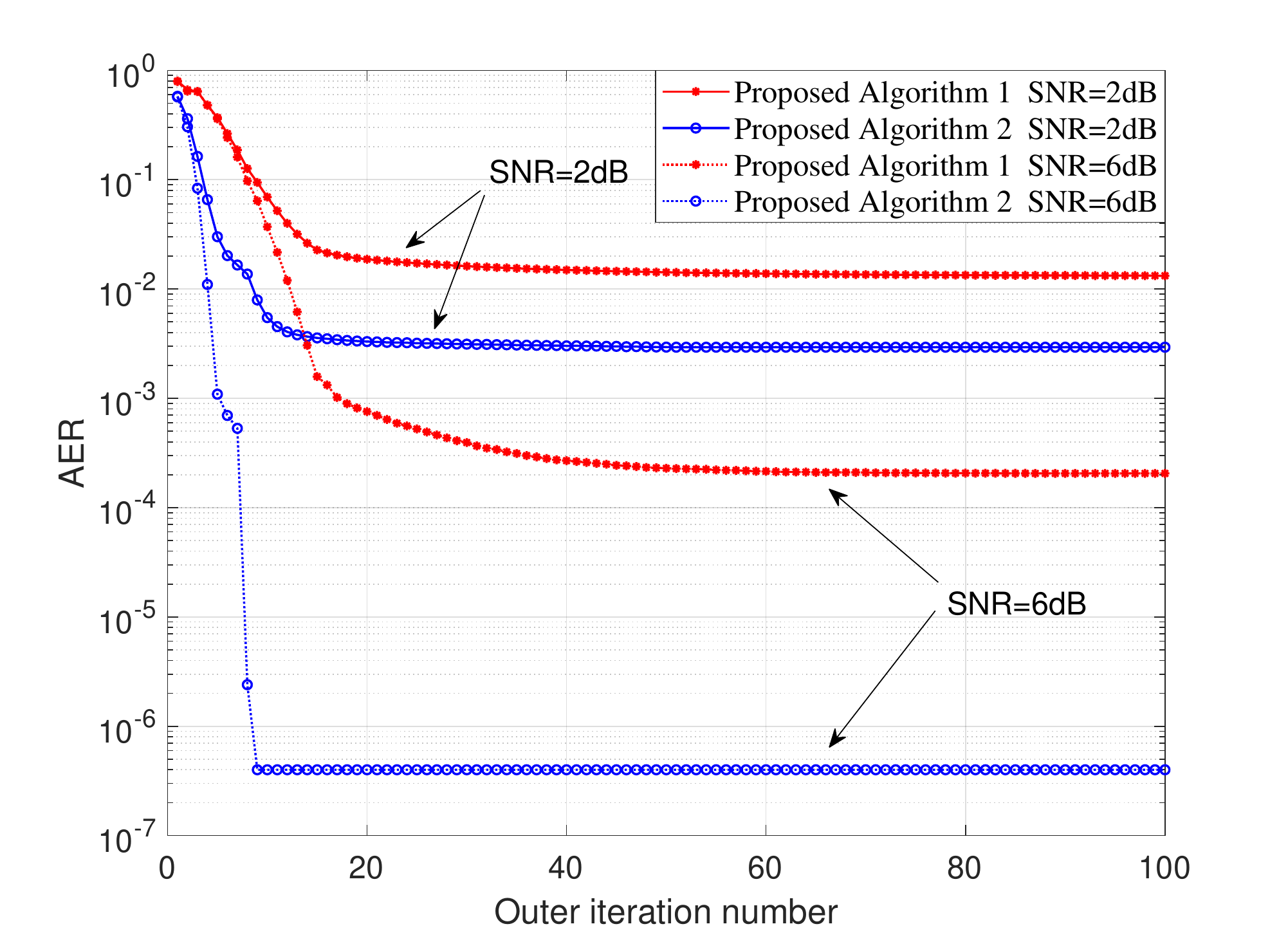}
	\caption{ Convergence of AER. }\label{fig:AERitr}
\end{figure}

\begin{figure}[t]
	\centering
	\includegraphics[scale=0.4]{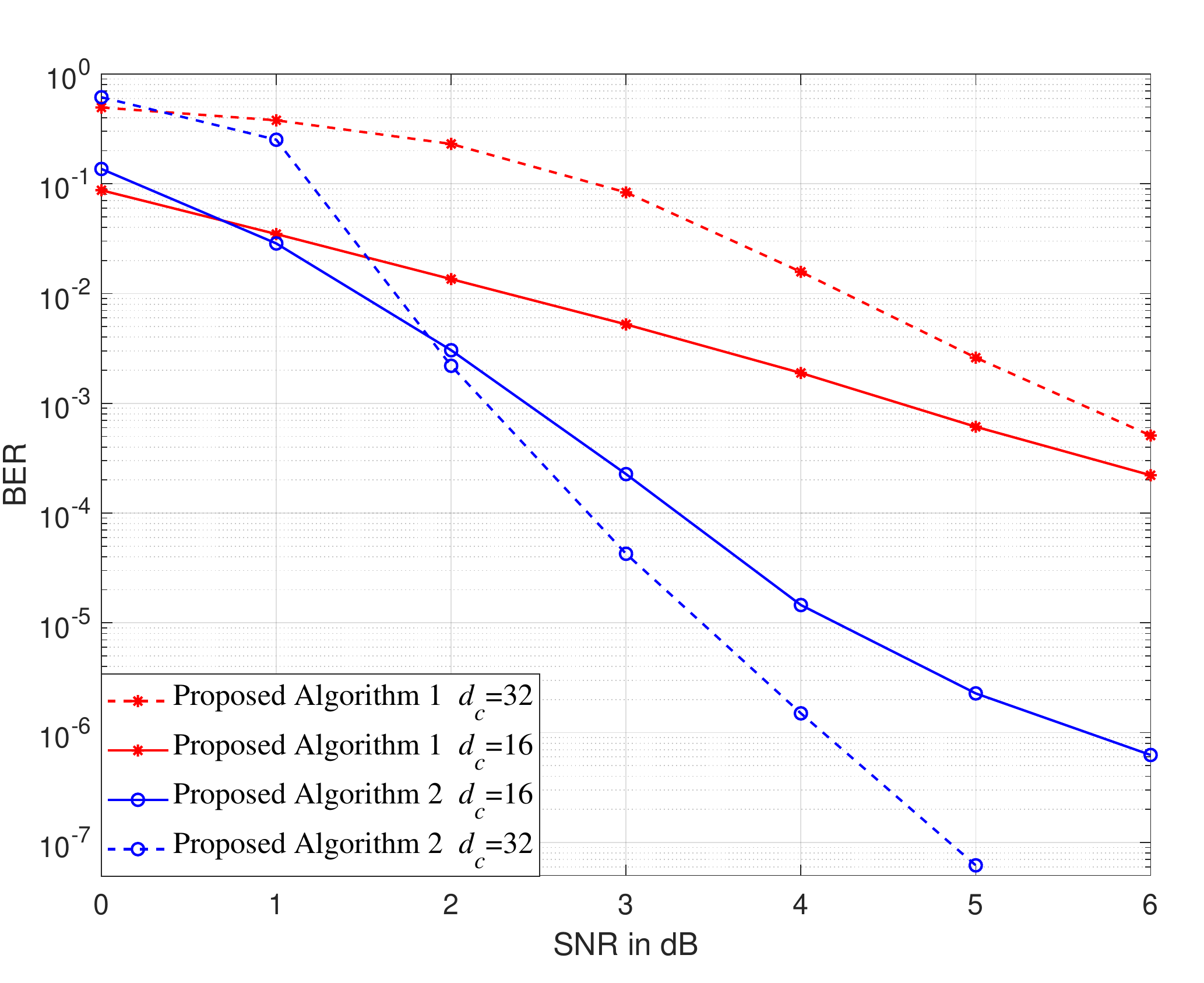}
	\caption{ BER performance comparison for different $d_c$. }\label{fig:BER16v32}
\end{figure}

\begin{figure}[t]
	\centering
	\includegraphics[scale=0.45]{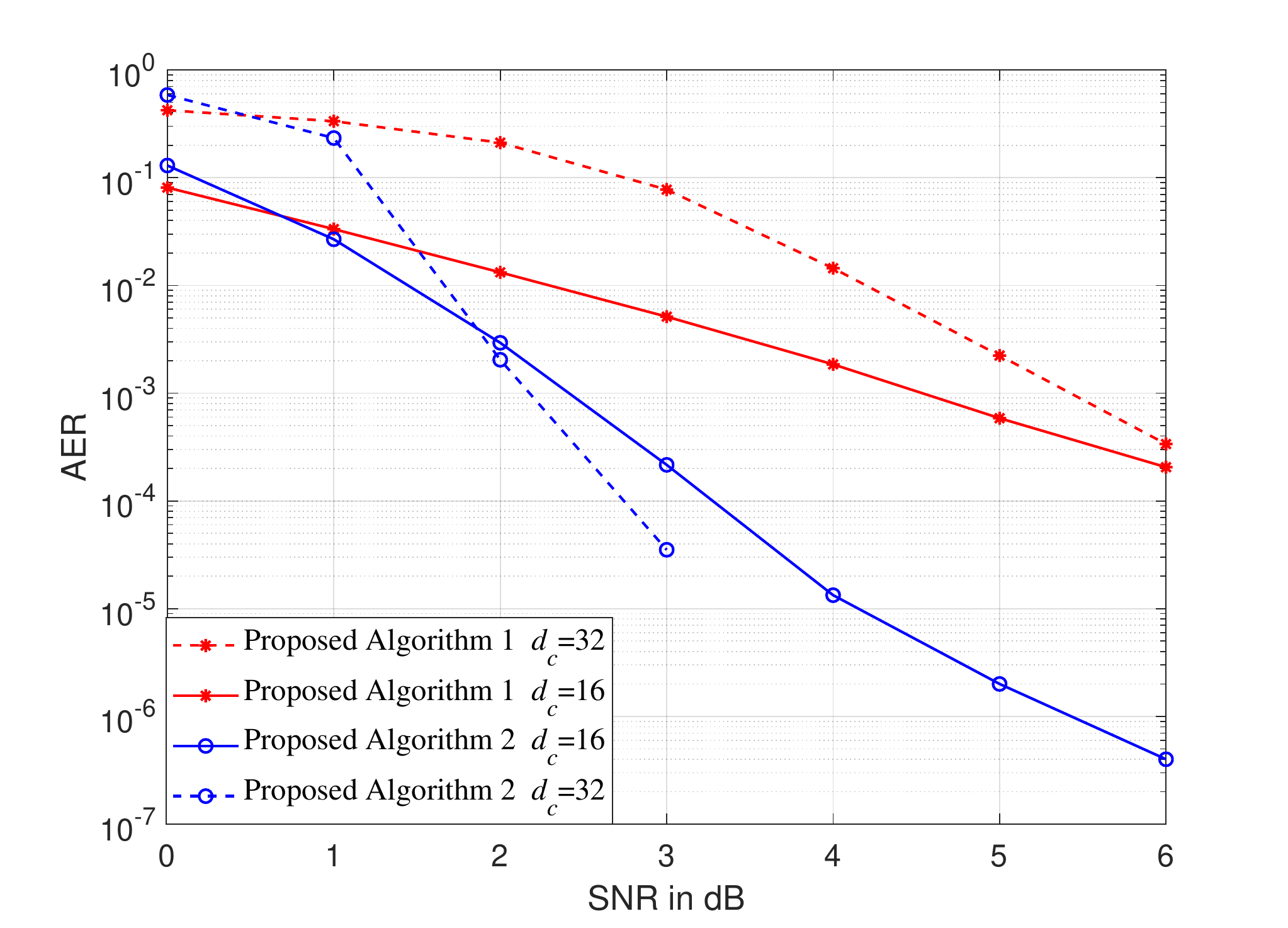}
	\caption{ AER performance comparison for different $d_c$. }\label{fig:AER16v32}
\end{figure}

\begin{figure}[t]
	\centering
	\includegraphics[width=3.5in]{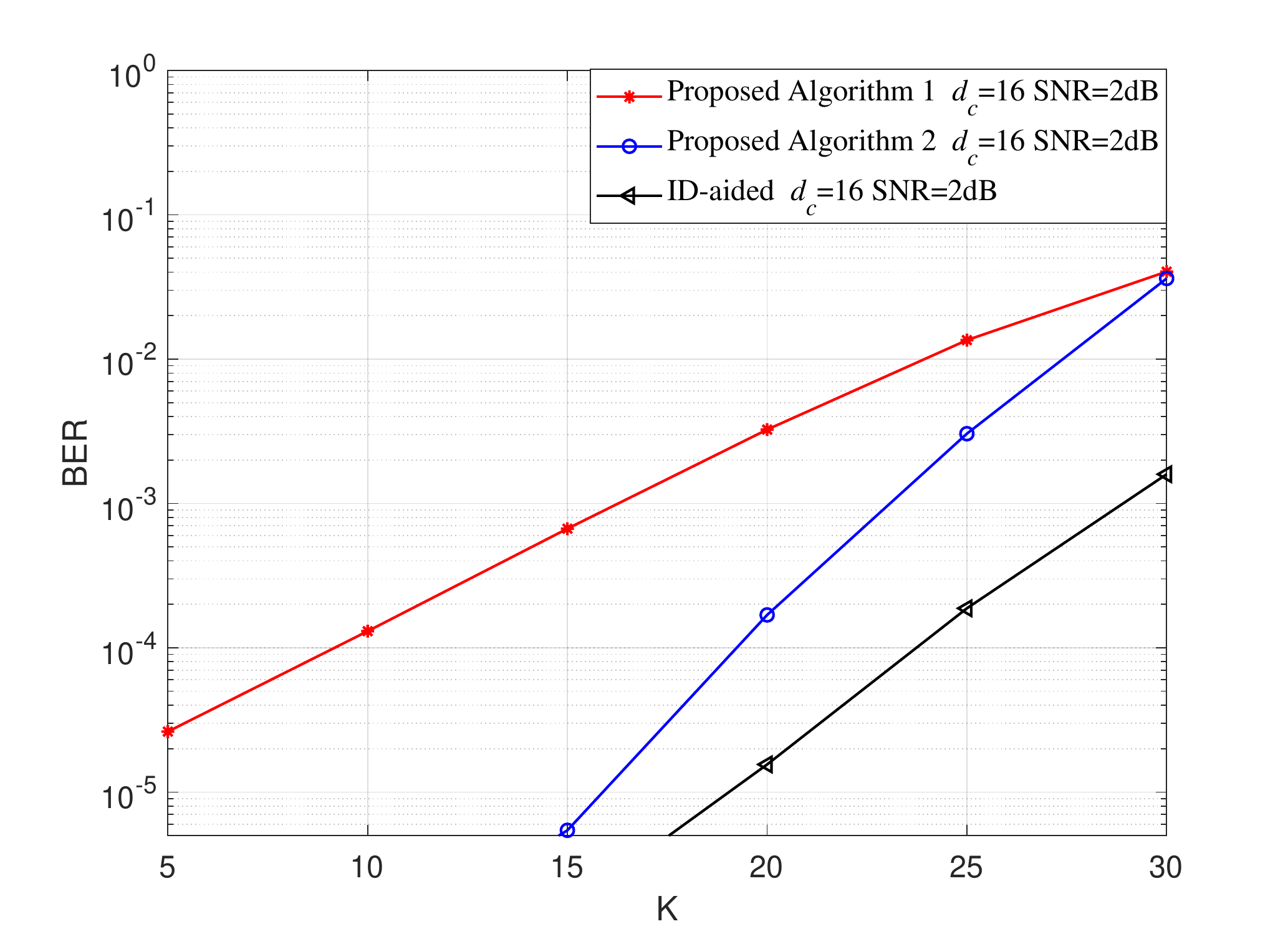}
	\caption{ BER performance versus active user number where $ L=40 $. }\label{fig:BERVarK}
\end{figure}

\begin{figure}[t]
	\centering
	\includegraphics[width=3.5in]{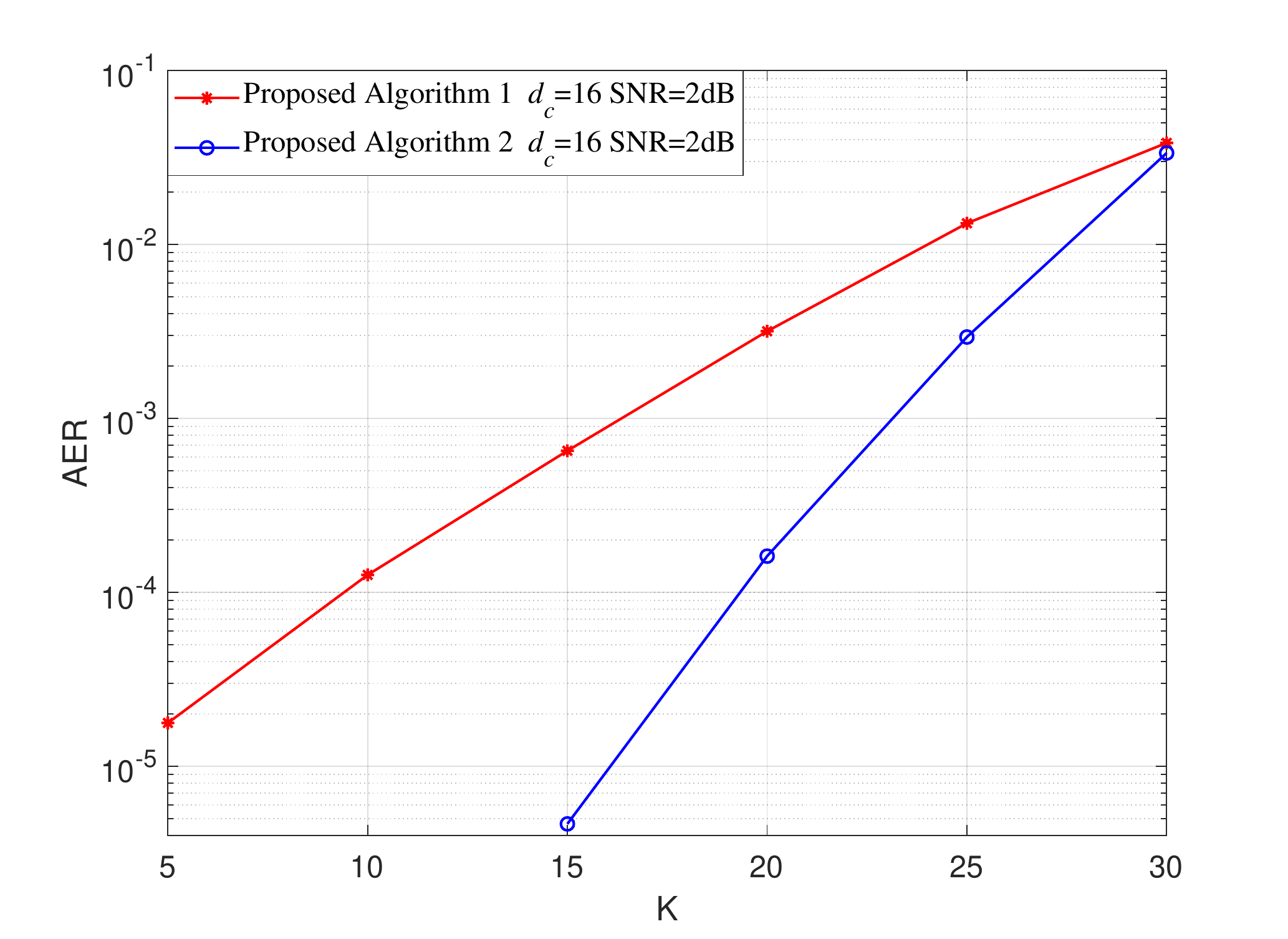}
	\caption{ AER performance versus active user number where $ L=40 $.  }\label{fig:AERVarK}
\end{figure}

Fig. \ref{fig:AER} compares the AER performance of different  algorithms with respect to SNRs. It can be seen that  {Algorithm \ref{alg:H-MP}} can outperform  {Algorithm \ref{alg:MF-MP}}  dramatically at relatively high SNRs. The BER and AER convergence of the proposed schemes are shown in Fig. \ref{fig:BERitr} and Fig. \ref{fig:AERitr}, respectively. We can see that the convergence rate of the receiver with  {Algorithm \ref{alg:H-MP}} is obviously quicker than that of  {Algorithm \ref{alg:MF-MP}}. The AER and BER of  {Algorithm \ref{alg:H-MP}}  falls rapidly within the first 10 iterations and it converges in about $ N_{O_{itr}}\!\!=10 $ for SNR = 6dB and about $ N_{O_{itr}}\!\!=20 $ for SNR = 2dB. Here we note that, the first 5 iterations are used for Algorithm \ref{preA2} to provide initial values in Algorithm \ref{alg:H-MP}. By comparison, Algorithm \ref{alg:MF-MP} converges slower, e.g., it requires about  $ N_{O_{itr}}\!\!=50 $ for SNR = 6dB and about $ N_{O_{itr}}\!\!=20 $ for SNR = 2dB.

With different  numbers of subcarriers occupied by each user, the BER and AER performance of the receiver with two algorithms are shown in Fig. \ref{fig:BER16v32} and Fig. \ref{fig:AER16v32}, respectively.  More subcarriers occupied by each user will lead to  larger frequency diversity gain, but stronger multi-user interference.
It can be observed that,  {Algorithm \ref{alg:MF-MP}} with $ d_c =32$ delivers worse performance than that with $ d_c =16$. This is because  {Algorithm \ref{alg:MF-MP}} has limited capability to handle the multi-user interference at the observation factors  and it is overwhelmed by multi-user interference, thereby leading to poor performance when  $ d_c =32$. In contrast,  {Algorithm \ref{alg:H-MP}} is able to handle the multi-user interference much more effectively. As we can see from Fig. \ref{fig:BER16v32} and Fig. \ref{fig:AER16v32} that, when SNR>2dB,  {Algorithm \ref{alg:H-MP}} with $ d_c =32$ performs considerably better than that with $ d_c =16$, i.e.,  {Algorithm \ref{alg:H-MP}} enjoys the diversity gain after mitigating the multi-user interference. It is noted that, in Fig. \ref{fig:AER16v32}, when SNR>3dB, all active users are identified correctly over $10^5$ trials.

Finally, we examine the BER and AER performance of the proposed algorithms by varying the number of active users, and the results are shown in Fig.  \ref{fig:BERVarK} and Fig. \ref{fig:AERVarK}, respectively, where the ID-aided scheme is also included for reference. It can be seen that, with the decrease of active user number $ K $, the multi-user interference is alleviated, which leads to better BER and AER performance for the proposed algorithms. As we can see, with the decrease of $ K $, the performance of  {Algorithm \ref{alg:H-MP}} approaches  the  ID-aided scheme closely.

\section{Conclusion}
In this paper, we have investigated the receiver design for grant-free LDS-OFDM, where pilot signals are not used to improve the transmission efficiency in IoT applications. The receiver has been implemented by solving a formulated structured signal estimation problem, where the structures of the equivalent channel matrix $ \textbf{H} $ and signal matrix $ \textbf{X} $ are fully exploited. Efficient hybrid message passing algorithms have been developed to solve the structured signal estimation problem. Simulation results have verified the effectiveness of the proposed algorithms.

\bibliographystyle{IEEEtran}
\bibliography{Bib}

\end{document}